\documentclass[aip, reprint]{revtex4-2}

\usepackage{graphicx}
\usepackage{dcolumn}
\usepackage{bm}
\usepackage{color}
\usepackage{scrextend}		
\usepackage{url}
\usepackage{epstopdf}
\usepackage{hyperref}

\begin{document}

\title{Dynamic Fingerprints of Synthetic Antiferromagnet Nanostructures with Interfacial Dzyaloshinskii-Moriya Interaction}

\author{Martin Lonsky}
\email{mlonsky@physik.uni-frankfurt.de}
\affiliation{Materials Research Laboratory and Department of Materials Science and Engineering, University of Illinois at Urbana-Champaign, Urbana, Illinois 61801, USA}
\affiliation{Institute of Physics, Goethe University Frankfurt, 60438 Frankfurt, Germany}
\author{Axel Hoffmann}
\affiliation{Materials Research Laboratory and Department of Materials Science and Engineering, University of Illinois at Urbana-Champaign, Urbana, Illinois 61801, USA}

\date{\today}

\begin{abstract}
Synthetic antiferromagnet (SAF) nanostructures with interfacial Dzyaloshinskii-Moriya interaction can host topologically distinct spin textures such as skyrmions and thus are regarded as promising candidates for both spintronics and magnonics applications. Here, we present comprehensive micromagnetic simulations of such material systems and discuss the rich phase diagrams that contain various types of magnetic configurations. Aside from the static properties, we further discuss the resonant excitations of the calculated magnetic states which include individual skyrmions and skyrmioniums. Finally, the internal modes of SAF skyrmion clusters are studied and discussed in the context of magnetic sensing applications based on the dynamic fingerprint in broadband ferromagnetic resonance measurements.   
\end{abstract}

\maketitle

\section{Introduction} \label{intro}
Synthetic antiferromagnets (SAFs) have attracted extensive attention over the past years owing to their highly tunable static and dynamic magnetic properties, reduced stray fields, as well as potentially fast current-driven dynamics of magnetic solitons such as domain walls or skyrmions.\cite{Duine2018, Waring2020, Dohi2019} Numerous possible applications of SAFs have been proposed in two major subfields of magnetism, spintronics\cite{Duine2018} and magnonics.\cite{Dai2021, Sorokin2020}

In general, a SAF is composed of thin ferromagnetic (FM) layers that are coupled antiferromagnetically through a nonmagnetic spacer layer by means of the Ruderman-Kittel-Kasuya-Yosida (RKKY) interaction.\cite{Parkin1991, Bruno1991} Through the introduction of additional heavy-metal layers (e.g., platinum), it becomes possible to engineer SAFs such that an interfacial Dzyaloshinskii-Moriya interaction (DMI) is present in addition to the conventional symmetric exchange interaction.\cite{Dzyaloshinskii1958, Moriya1960} Recently, it has been demonstrated experimentally that SAFs can host magnetic skyrmions even at room temperature.\cite{Legrand2020, Dohi2019, Chen2020, Juge_2021}    

For the case of FM materials it was shown that confined skyrmions in lithographically-defined nanostructures such as nanodisks, nanodots, nanowires or nanostrips are of particular interest.\cite{Mathur2019} Such geometries provide unique opportunities for the controlled nucleation, propagation or excitation of topologically distinct spin textures. In contrast to their ferromagnetic counterparts, literature on SAF nanostructures with interfacial DMI that can potentially host skyrmions or other chiral spin textures still remains scarce, despite of their interesting and partially advantageous properties. For instance, in previous studies we have shown that skyrmions in SAF nanodisks exhibit internal resonance modes whose properties (e.g., their resonance frequencies) can be tuned by various parameters such as the interlayer exchange coupling or the magnetic anisotropy.\cite{Lonsky2020, Lonsky2020a} The observed breathing modes can then be exploited for driving skyrmions across SAF nanostructures with significantly higher velocities than in ferromagnets.\cite{Qiu2021}    

Here, we present extensive micromagnetic simulations of such systems. In detail, we discuss the various (meta)stable states observed in SAF nanodisks with DMI on the basis of phase diagrams that we determined for different initial magnetic configurations and interlayer antiferromagnetic coupling strength. This is followed by an analysis of the resonant spin dynamics for the obtained states, which include complex objects such as skyrmioniums, deformed skyrmions or skyrmion clusters. We note that similar systematic studies of magnetic properties have been reported on chiral magnetic textures in ferromagnetic\cite{Winkler_2021} and bulk antiferromagnetic nanostructures.\cite{Ji2021} In analogy to previous works, we will demonstrate that dynamic fingerprints in the resonance mode spectra of metastable magnetization configurations contain valuable information and merit a systematic investigation.\cite{Gliga2013} In particular, this is true for synthetic antiferromagnets, where the conventional detection of magnetic states is complicated due to the fully compensated magnetic moments and therefore a vanishing stray field. 
             
\section{Micromagnetic Modeling}
We performed extensive micromagnetic simulations of FM and SAF nanostructures by using the Object Oriented MicroMagnetic Framework (\textsc{oommf}) code.\cite{Donahue1999, Rohart2013} This software package carries out a numerical time integration of the Landau-Lifshitz-Gilbert (LLG) equation for the local magnetization:
\begin{equation} \label{eq:LLG}
\frac{\mathrm{d}\textbf{m}}{\mathrm{d}t}=-|\gamma_{0}|\textbf{m}\times \textbf{H}_{\mathrm{eff}}+\alpha \textbf{m}\times \frac{\mathrm{d}\textbf{m}}{\mathrm{d}t}.
\end{equation} 
Here, $\gamma_{0}$ denotes the gyromagnetic ratio, $\alpha$ is the Gilbert damping parameter, $\textbf{m}$ is the unit vector of the magnetization, and $\textbf{H}_{\mathrm{eff}}$ is the effective magnetic field that is proportional to the derivative of the total micromagnetic energy $E$ with respect to the magnetization. In our model, we assume that $E$ is given by the following expression: 
\begin{equation} \label{eq:Utot}
E = E_{\mathrm{Ex}} + E_{\mathrm{Z}} + E_{\mathrm{Dem}} + E_{\mathrm{Anis}} + E_{\mathrm{DMI}}.
\end{equation}
In Eq.\ (\ref{eq:Utot}), $E_{\mathrm{Z}}$ denotes the Zeeman energy, $E_{\mathrm{Ex}}$ the isotropic exchange interaction characterized by the exchange stiffness $A$, $E_{\mathrm{Dem}}$ the demagnetization energy, $E_{\mathrm{Anis}}$ the uniaxial anisotropy energy, and $E_{\mathrm{DMI}}$ the interfacial Dzyaloshinskii-Moriya interaction (DMI) that is given by
\begin{equation}
E_{\mathrm{DMI}}=D\left[m_{z}\left(\nabla \cdot \textbf{m}\right)-\left(\textbf{m} \cdot \nabla \right)m_{z} \right],
\end{equation} 
where $D$ is the DMI constant specifying the strength of the interaction.\cite{Dzyaloshinskii1958, Moriya1960}

For the specific case of SAF nanostructures, an additional energy term $E_{\mathrm{RKKY}}$ is required to account for the antiferromagnetic interlayer exchange interaction between the two ultrathin FM layers that are coupled via a nonmagnetic spacer through a Ruderman-Kittel-Kasuya-Yosida (RKKY) interaction which is described by:\cite{Parkin1991, Bruno1991, Lonsky2020}    
\begin{equation}\label{eq:RKKY}
E_{\mathrm{RKKY}}=\frac{\sigma}{t_{\mathrm{NM}}}\left(1-\textbf{m}_{\mathrm{t}}\textbf{m}_{\mathrm{b}}\right).
\end{equation}
Here, $\sigma$ corresponds to the surface exchange coefficient and $t_{\mathrm{NM}}$ is the thickness of the nonmagnetic spacer layer. Moreover, $\textbf{m}_{\mathrm{t}}$ and $\textbf{m}_{\mathrm{b}}$ are the unit vectors of magnetization for the top and bottom layer, respectively.   

The main geometry considered in our numerical modeling is that of a SAF nanodisk with diameter $d=100\,$nm. More specifically, it consists of two FM films that are separated by a nonmagnetic layer ($t = 1\,$nm for each of the three layers). The assumed FM material corresponds to a typical thin-film perpendicular anisotropy system with an exchange stiffness $A=8\,$pJ/m, while the perpendicular anisotropy constant is varied between $K_{\mathrm{u}}=0.4\,$MJ/m$^3$ and $K_{\mathrm{u}}=1.4\,$MJ/m$^3$, and the saturation magnetization between $\mu_{0}M_{\mathrm{s}}=0.6\,$T and $\mu_{0}M_{\mathrm{s}}=1.8\,$T. The interlayer RKKY-coupling strength is set to $\sigma=-3\times 10^{-4}\,$J/m$^2$ initially, but subsequent simulations also assume either $\sigma=-3\times 10^{-3}\,$J/m$^2$ or $\sigma=-3\times 10^{-5}\,$J/m$^2$. The magnitude of the interfacial DMI is $D=3\,$mJ/m$^2$ for both ferromagnetic layers. In addition to the SAF, we have also modeled a single $1\,$nm thin FM nanodisk with the simulation parameters being identical to one of the FM layers in the SAF nanodisk. This allows for a direct comparison between the properties of FM and SAF nanodisks with DMI. Lastly, we have considered a square-shaped SAF nanostructure with $100\,$nm$\times 100\,$nm lateral size and otherwise identical parameters. In all of our simulations we have defined a mesh consisting of finite difference cells of size $1\,$nm$\times 1\,$nm$\times 1\,$nm. 

Starting from different initial magnetization states (see details further below), the (meta)stable states in all nanostructures have been determined by relaxing the system for at least $20\,$ns with a high damping constant of $\alpha = 0.5$, whereby the time evolution of the magnetization was monitored to confirm that an equilibrium state has been reached. No dc magnetic fields have been applied.    
Subsequently, the resonant dynamics of selected magnetic nanostructures was studied after exposing them to a spatially uniform time-varying ac magnetic field pulse $H_{\mathrm{ac}}=H_{0}\sin(2\pi f t)/(2\pi f t)$ along the $z$-axis, where $\mu_{0}H_{0}=0.5\,$mT is the amplitude and $f=200\,$GHz the cutoff frequency. For these simulations, the damping parameter has been chosen as $\alpha=5\times 10^{-3}$ to ensure a satisfactory frequency resolution of the excited modes. The dynamics is simulated for at least $6\,$ns with data taken every $2\,$ps. Eventually, the power spectral density (PSD) of the spatially-averaged $z$-component of the magnetization $\langle m_{z}\rangle (t)$ is calculated by using a fast Fourier transform (FFT) algorithm.

\section{Results and Discussion} \label{RESULTS}
\subsection{Magnetic States in a FM Nanodisk with DMI}
\begin{figure}
\centering
\includegraphics[width=8.5 cm]{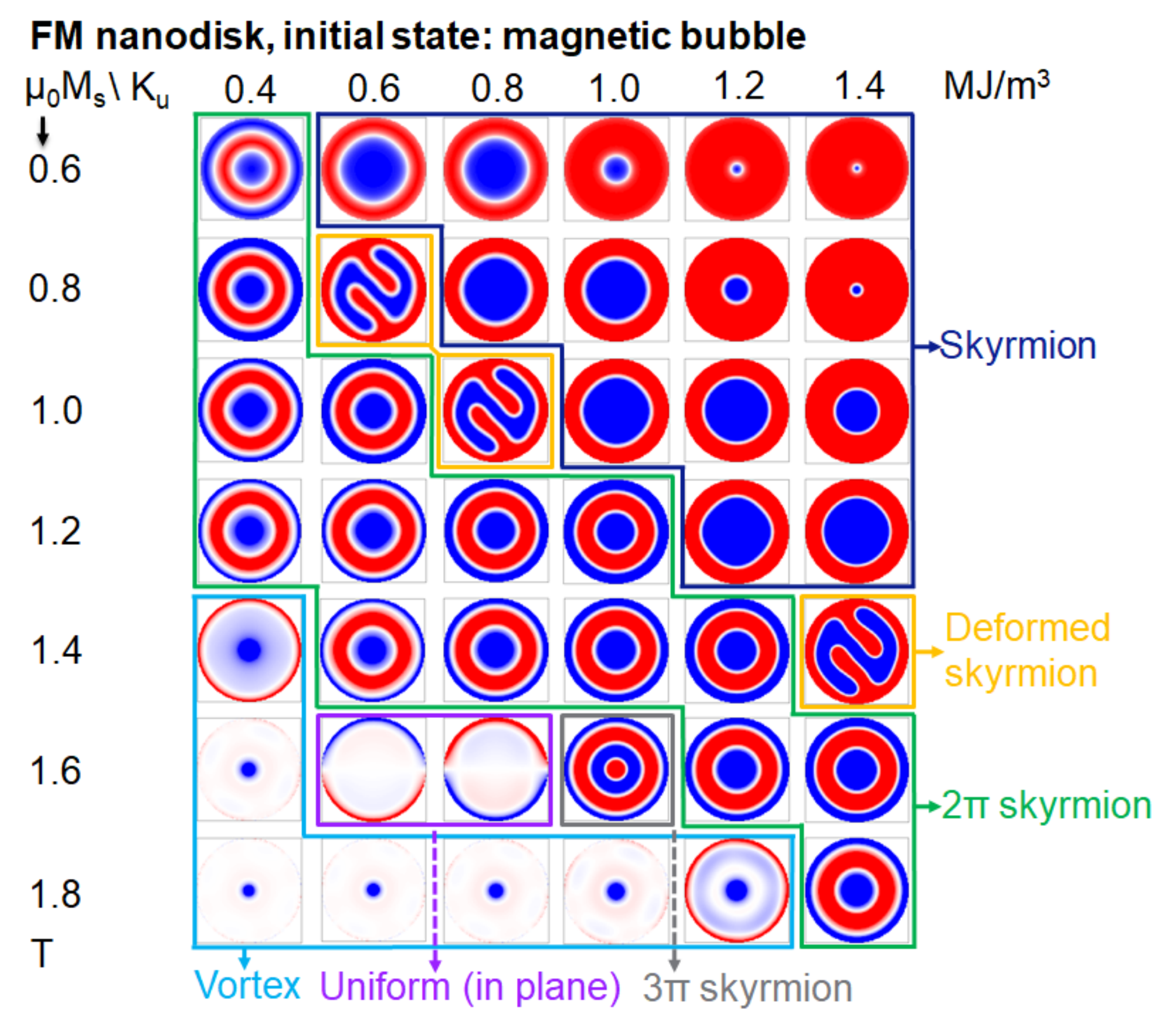}
\caption{Calculated magnetic configuration of a FM nanodisk with diameter $d=100$\,nm, thickness $t=1$\,nm, varying saturation magnetization $M_{\mathrm{s}}$ and uniaxial anisotropy constant $K_{\mathrm{u}}$. Colors indicate the out-of-plane component ($m_{z}$) of the normalized magnetic moment $m$; blue: $m_{z}=-1$, white: $m_{z}=0$, red: $m_{z}=+1$.} 
\label{FMDISK}%
\end{figure}

We begin the discussion by considering the calculated metastable states in a FM nanodisk with varying magnetic parameters, which we will then compare to a SAF nanodisk of equal lateral size. Figure \ref{FMDISK} illustrates the phase diagram of the simulated minimized energy states of the FM nanodisk, whereby the saturation magnetization $\mu_{0}M_{\mathrm{s}}$ is assumed to lie between $0.6$\,T and $1.8$\,T, and the uniaxial anisotropy constant $K_{\mathrm{u}}$ ranges from $0.4$\,$\mathrm{MJ/m^{3}}$ to $1.4$\,$\mathrm{MJ/m^{3}}$. Here, the different colors indicate the out-of-plane component ($m_{z}$) of the normalized magnetic moment $m$: blue corresponds to $m_{z}=-1$, white to $m_{z}=0$, and red to $m_{z}=+1$. As indicated in the figure, the initial magnetic state prior to relaxation was that of a small magnetic bubble (diameter: $25\,$nm) with its magnetization pointing along the negative $z$-direction ($m_{z}=-1$), while the magnetization in the remaining part of the nanodisk points along the positive $z$-direction ($m_{z}=+1$).

We observe minimized energy states that include a skyrmion, skyrmionium (more precisely, a $2\pi$ and $3\pi$ skyrmion), S-shaped band-domain with a net topological charge $n=\left(1/4\pi\right) \int \textbf{m}\cdot \left(\partial_{x}\textbf{m}\times \partial_{y}\textbf{m}\right)\,\mathrm{d}x\,\mathrm{d}y=1$ equivalent to a skyrmion,\cite{Siracusano_2016, Beg_2015} and a magnetic vortex state. Starting from the top right corner of the diagram, the diameter of the skyrmion becomes larger for (i) a decreasing magnetic uniaxial anisotropy and (ii) an increasing saturation magnetization $\mu_{0}M_{\mathrm{s}}$. At a certain threshold level (that is, in the case of a small enough anisotropy or sufficiently large saturation magnetization) the skyrmion diameter approaches the total disk diameter of $d=100\,$nm, which results in the skyrmion solution becoming increasingly less favorable. Therefore, the skyrmion transforms into an S-shaped skyrmion-like state (e.g., for $\mu_{0}M_{\mathrm{s}}=0.8\,$T and $K_{\mathrm{u}}=0.6\,$MJ/m$^3$) and/or a skyrmionium (e.g., for $\mu_{0}M_{\mathrm{s}}=1.2\,$T and $K_{\mathrm{u}}=1.0\,$MJ/m$^3$) to minimize the total micromagnetic energy $E$, see also Eq.\ (\ref{eq:Utot}). As can be seen in Fig.\ \ref{FMDISK}, the transition regime of $\mu_{0}M_{\mathrm{s}}$ and $K_{\mathrm{u}}$ values for which the skyrmionium states occur, is fairly broad. As indicated in the lower left corner of the diagram, a magnetic vortex state becomes energetically more favorable toward even lower out-of-plane anisotropy and higher saturation magnetization. To summarize, it is possible to tune the magnetic configuration of a FM nanodisk systematically in our simulations. We emphasize that there are many more approaches to do so, such as the application of dc magnetic fields,\cite{Kim2014} varying the DMI strength\cite{Winkler_2021} or exploiting magnetoelastic effects.\cite{Hu2020}  

In the following, we address the question how magnetic compensation may affect the metastable states in a nanodisk. The compensation is realized by adding a second FM layer that is coupled to the first FM via a RKKY-like interaction through a nonmagnetic spacer layer. Furthermore, we will also analyze the response to external ac magnetic fields of such SAF nanostructures. For the case of FM nanodisks, it is well-established that the distinct magnetic states exhibit characteristic resonance modes, see the discussion in previous works for skyrmions and skyrmioniums\cite{Kim2014, Song2019} as well as for magnetic vortices\cite{Guslienko_2002, Choe2004}. In detail, these studies have indicated that broadband ferromagnetic resonance measurements may constitute a powerful tool to detect different types of magnetic states in nanostructures. However, for the case of SAFs, such investigations thus far have been limited to skyrmions\cite{Lonsky2020} and skyrmion strings.\cite{Barker_2021}

\subsection{Magnetic States in a SAF nanodisk with DMI}
\begin{figure*}
\centering
\includegraphics[width=17.5 cm]{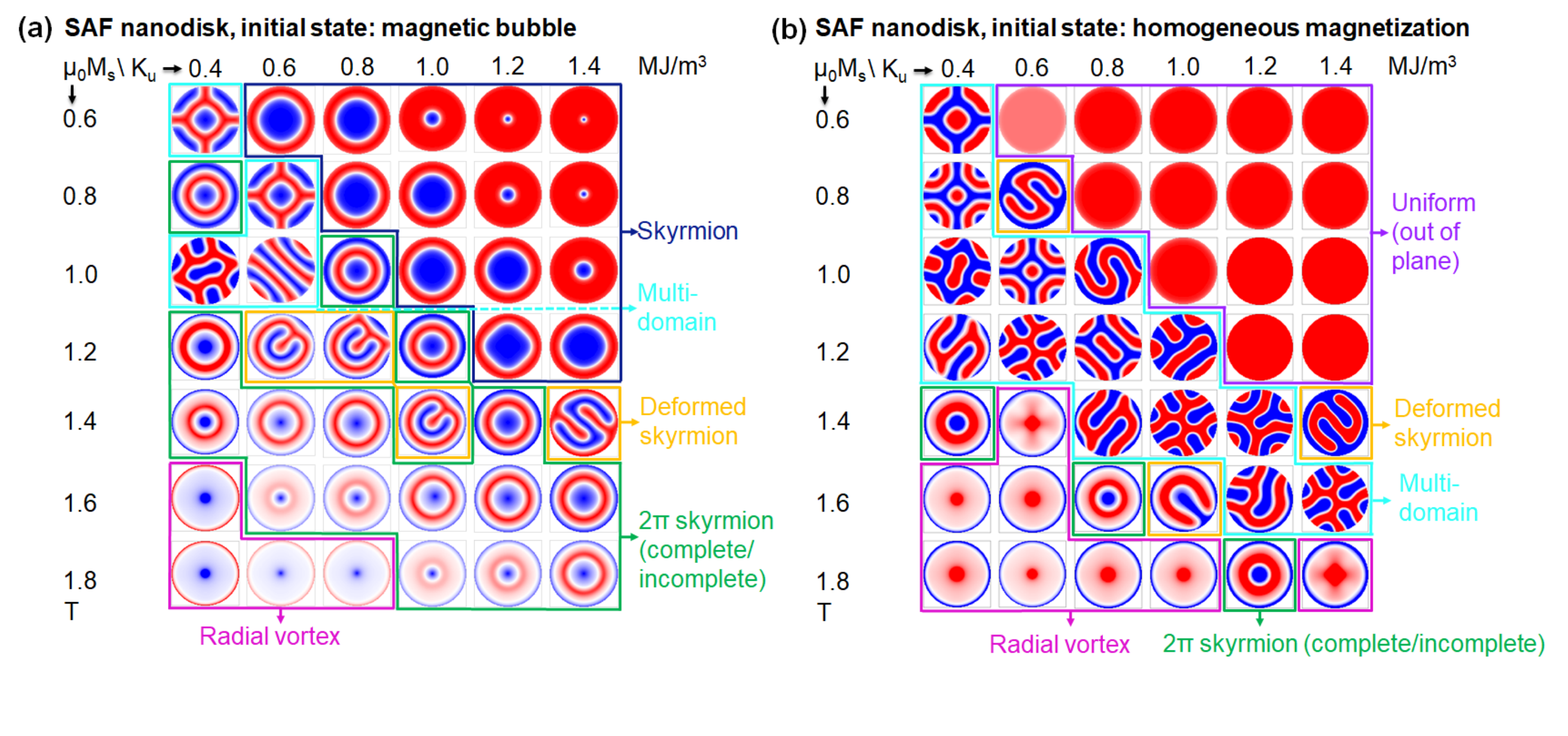}
\caption{Phase diagrams of SAF nanodisks ($d=100$\,nm) with interfacial DMI, determined for varying saturation magnetization $M_{\mathrm{s}}$ and uniaxial anisotropy constant $K_{\mathrm{u}}$. Interlayer RKKY coupling is of moderate strength, $\sigma=-3\times 10^{-4}\,$J/m$^2$. Initial magnetization state for the numerical modeling has been assumed as (a) a fully magnetized state along the $z$-axis ($m_{z}=+1$ for one FM layer, $m_{z}=-1$ for the other) with a small bubble that has a $m_{z}$ of opposite sign and (b) a homogeneously magnetized state. Shown is only the top FM layer.} 
\label{SAFDISK_INITCOND}%
\end{figure*}

Initially, we have calculated the minimized energy states in a SAF nanodisk with $d=100\,$nm for an antiferromagnetic coupling strength of $\sigma=-1\times 10^{-4}\,$J/m$^2$ at zero external magnetic field. Figure \ref{SAFDISK_INITCOND} provides an overview of magnetic configurations for two types of initial magnetization states. Note that only the magnetization state of the top FM layer is depicted here, but we have verified that the bottom FM layer, which is coupled antiferromagnetically to the top layer, exhibits exactly the same state with opposite magnetization direction (unless stated otherwise). In the first case, Fig.\ \ref{SAFDISK_INITCOND}(a) illustrates the relaxed states calculated after initializing the magnetic configuration with two magnetic bubbles ($d=25\,$nm) where $m_z$ is of opposite sign in the two FM layers. In the second case, it is evident from \ref{SAFDISK_INITCOND}(b) that a majority of the states are significantly different when the SAF nanodisk system is initialized in a fully magnetized state (constant $m_{z}=+1$ and $m_{z}=-1$ in the top and bottom FM layers, respectively). This discrepancy between the states depicted in Fig.\ \ref{SAFDISK_INITCOND}(a) and (b) can be attributed to the existence of multiple local energy minima. Depending on the choice of the initial magnetization configuration, the calculated metastable state may or may not correspond to a chiral magnetic texture. In fact, when using the homogeneously magnetized state as the initial configuration, only rarely do skyrmionium-like objects appear as the minimized energy state, while skyrmions are not present in the diagram at all. Consequently, a general implication of our results depicted in Fig.\ \ref{SAFDISK_INITCOND} is the importance of the initial magnetic state in micromagnetic modeling, which in numerous publications is not even mentioned explicitly. In addition, we note that a further complication is given by the quadrilateral nature of the simulation cells, which do not exactly reproduce the circular boundary of a nanodisk. A cylindrical mesh may in principle constitute a more natural way to model such a geometry, but the use of a sufficiently large number of quadrilateral cells still represents an alternative valid approach.\cite{Castell_Queralt_2022} While we did not observe a dependence of our results on the simulation cell size, we emphasize that selected states such as the multi-domain configurations in Fig.\ \ref{SAFDISK_INITCOND} may be related to the existence of several local minima of similar energy and hence depend on the initial state.    

We now turn to the comparison of FM and SAF nanodisks. At first glance, Fig.\ \ref{SAFDISK_INITCOND}(a) strongly resembles the case of a FM nanodisk as shown in Fig.\ \ref{FMDISK}. However, there are subtle differences, such as the occurrence of a C-shaped skyrmion state and multi-domain (or spin spiral) states, as well the non-existence of magnetic vortex configurations. The number of states containing a skyrmionium (i.e., a $2\pi$ skyrmion) is reduced, and there is no $3\pi$ skyrmion observed in the SAF nanodisk. Furthermore, in the top left corner of the phase diagram we observe an interesting multi-domain state which exhibits a regular skyrmion in its core. When compared to regular skyrmionium states, this configuration can be regarded as an incomplete skyrmionium. 
In contrast to the FM nanodisk, we notice that instead of the conventional magnetic vortex states there exists a different type of configuration in the bottom left corner -- a so-called magnetic radial vortex.\cite{Siracusano_2016, Yan_2008} Here, the magnetization points along the positive $z$-direction in the center of the disk. Further away from the center and as indicated by the lighter colors, the moments possess a strong in-plane component and point radially toward the disk edge. At the edge, $m_{z}$ is antiparallel to its counterpart in the center of the disk. Due to the changes in the uniaxial anisotropy constant $K_{\mathrm{u}}$, the aforementioned behavior can be explained by the occurence of a spin-reorientation transition in the system, although a significant out-of-plane magnetization still remains even for high $M_{\mathrm{s}}$ and low $K_{\mathrm{u}}$ in our chosen parameter range. This results in a non-integer topological charge between $0.5$ and $1$.\cite{Siracusano_2016} In a work by Karakas \textit{et al.}, such radial vortices are observed experimentally in Pt/CoFeB/Ti multilayers.\cite{Karakas_2018} Our simulations suggest that such magnetic radial vortices are energetically favorable in selected SAF nanodisk structures as well.  

For the case of a fully magnetized initial state, see Fig.\ \ref{SAFDISK_INITCOND}(b), in the top right corner of the phase diagram we do not observe any skyrmion states. There are many more multi-domain states in the phase diagram, mostly along a diagonal from the top left to the bottom right corner. In contrast to the absence of magnetic skyrmions, we still find a few metastable skyrmionium states. We note that for other initial conditions than those discussed in this paper, such as those of antiferromagnetically-coupled magnetic vortices in both FM layers, one obtains a slightly different phase diagram again. As mentioned above, this can be ascribed to the existence of multiple local energy minimum states. In essence, depending on the initial conditions the simulation gets trapped in one of these local minima. In practice, one could switch between adjacent local minima by using ac electric current or magnetic field pulses, see for instance Ref.\ \cite{Lin_2013}

Our results can be compared to another comprehensive computational study on FM nanodisks by Winkler \textit{et al.}\cite{Winkler_2021} In their work, only one single C-shaped skyrmion has also been found in FM nanodisks for a very specific set of parameters. As can be seen in Fig.\ \ref{FMDISK}, we do not observe this subtype of deformed skyrmions in the case of our simulated FM nanodisk, whereas we do observe a few such states in the SAF. It is well known for magnetic vortex nucleation that C-shaped and S-shaped intermediate structures are very close in their energy.\cite{Guslienko_2001} Due to the relatively large step size of variations of saturation magnetization and magnetic anisotropy in our simulations, we may not be able to detect all C-shaped and S-shaped states. Therefore, it is not surprising that the number of these configurations is different for the FM and SAF nanodisk phase diagrams. However, the occurrence of such C-shaped configurations still appears to be enhanced in SAFs. Finally, we emphasize that both the S-shaped and C-shaped skyrmions exhibit a topological charge of $1$ despite not possessing a circular symmetry like regular skyrmions would do.\cite{Siracusano_2016} 
In summary, the two most apparent consequences of magnetic compensation for the relaxed magnetic states are in our case (i) the additional presence of C-shaped skyrmions as well as multi-domain states and (ii) the change from conventional magnetic vortex states to magnetic radial vortices. Furthermore, we have shown more generally that SAF nanostructures, too, exhibit a complex set of metastable magnetic configurations, which may be studied in experiments and could potentially be exploited for device applications.

\begin{figure*}
\centering
\includegraphics[width=17.5 cm]{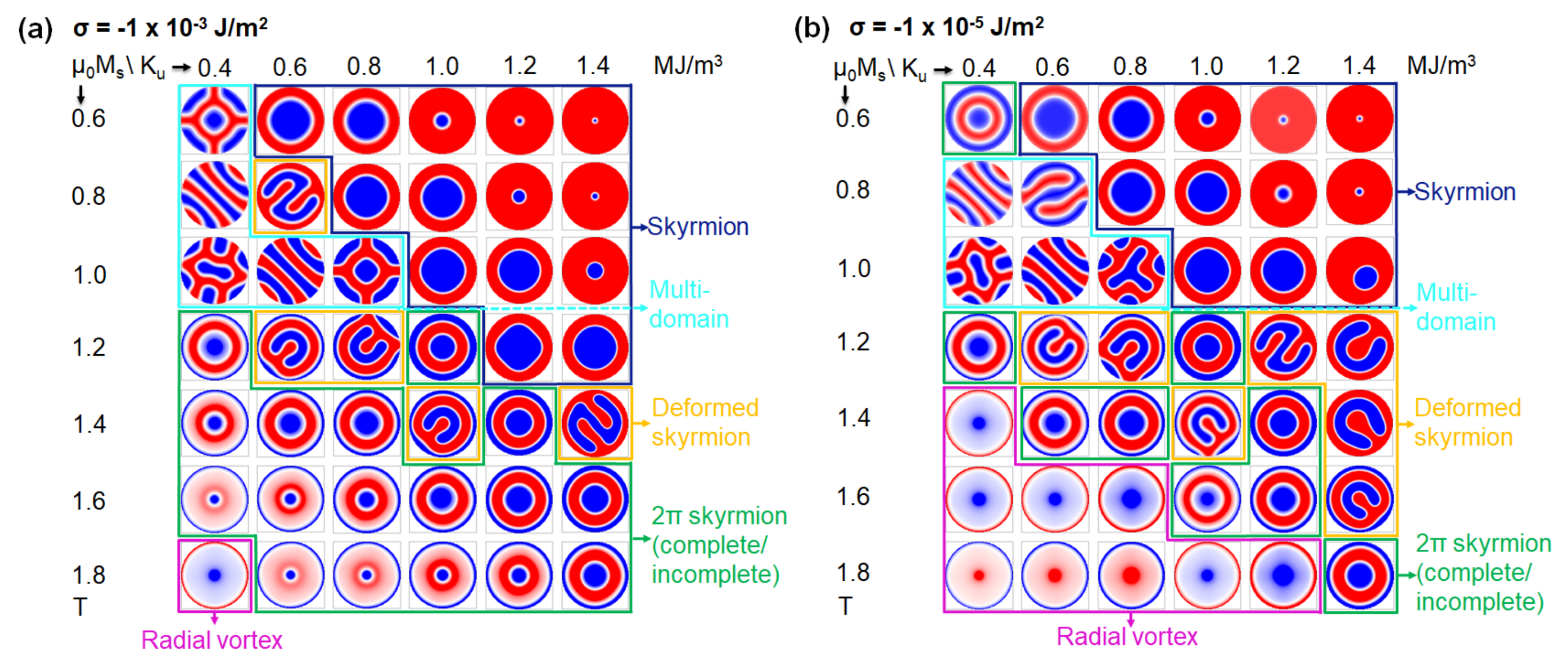}
\caption{Phase diagrams of SAF nanodisks ($d=100$\,nm) with interfacial DMI, determined for two different interlayer RKKY coupling strengths: (a) $\sigma=-3\times 10^{-3}\,$J/m$^2$ and (b) $\sigma=-3\times 10^{-5}\,$J/m$^2$. Shown is only the top FM layer.} 
\label{SAFDISK_COUPLING}%
\end{figure*}%

As shown in Ref.\ \cite{Lonsky2020}, there is a dependence of the static and dynamic properties on the interlayer coupling strength. Here, we further elaborate on this finding by discussing the phase diagram for two additional coupling strengths, namely $\sigma=-3\times 10^{-3}\,$J/m$^2$ and $\sigma=-3\times 10^{-5}\,$J/m$^2$. Depicted in Fig.\ \ref{SAFDISK_COUPLING} are phase diagrams for (a) a strong interlayer antiferromagnetic coupling of $\sigma=-3\times 10^{-3}\,$J/m$^2$ and (b) a comparably weak interlayer antiferromagnetic coupling strength of $\sigma=-3\times 10^{-5}\,$J/m$^2$. Interestingly, there are clearly more skyrmionium-like states in the bottom half of the phase diagram in the case of strong antiferromagnetic coupling, while for weaker coupling the magnetic moments preferably lie within the nanodisk plane, pointing towards the edges of the disk and thereby minimizing the demagnetization energy. Here, as discussed for Fig.\ \ref{SAFDISK_INITCOND}, in the case of radial vortices only the core and edge have a relatively strong $m_z$ component due to the weak uniaxial anisotropy of the system. For the strong coupling in Fig.\ \ref{SAFDISK_COUPLING}(a), the moments are forced to be oriented out-of-plane, and thus the system minimizes its stray field by exhibiting the configuration of a skyrmionium in each FM layer. It is conceivable that one could switch between the two aforementioned configurations in experiments by exploiting either voltage controlled magnetic anisotropy materials (VCMA)\cite{Rana_2019, Suwardy_2019} or by tuning the RKKY coupling strength continuously as a function of the electric voltage.\cite{Leon_2019}

It should also be emphasized that in the case of Fig.\ \ref{SAFDISK_COUPLING}(b), the three magnetic radial vortex states depicted in the bottom row with $\mu_{0}M_{\mathrm{s}}=1.8$\,T and $K_{\mathrm{u}}=0.4$, $0.6$ and $0.8$\,MJ/m$^3$ do not show an antiferromagnetic arrangement of the cores. This is due to the weak antiferromagnetic coupling strength and the dominance of dipolar coupling. It is known from previous studies that complex domain structures in SAFs may not be necessarily coherent in the coupled layers.\cite{Hellwig_2007} Overall, the most noticeable differences between the two distinct coupling strengths can be found in the bottom left corner of the phase diagram. Aside from that, we also emphasize the occurrence of multiple C-shaped skyrmions for both coupling strengths, with the highest number being observed for $\sigma=-3\times 10^{-5}\,$J/m$^2$. Following the analysis of static magnetic properties, we will in the next section discuss the resonant dynamics of the various magnetic configurations in a SAF nanodisk with interfacial DMI.     

\subsection{Magnetic Excitation Spectrum of a SAF Nanodisk with DMI}
\begin{figure}
\centering
\includegraphics[width=8.5 cm]{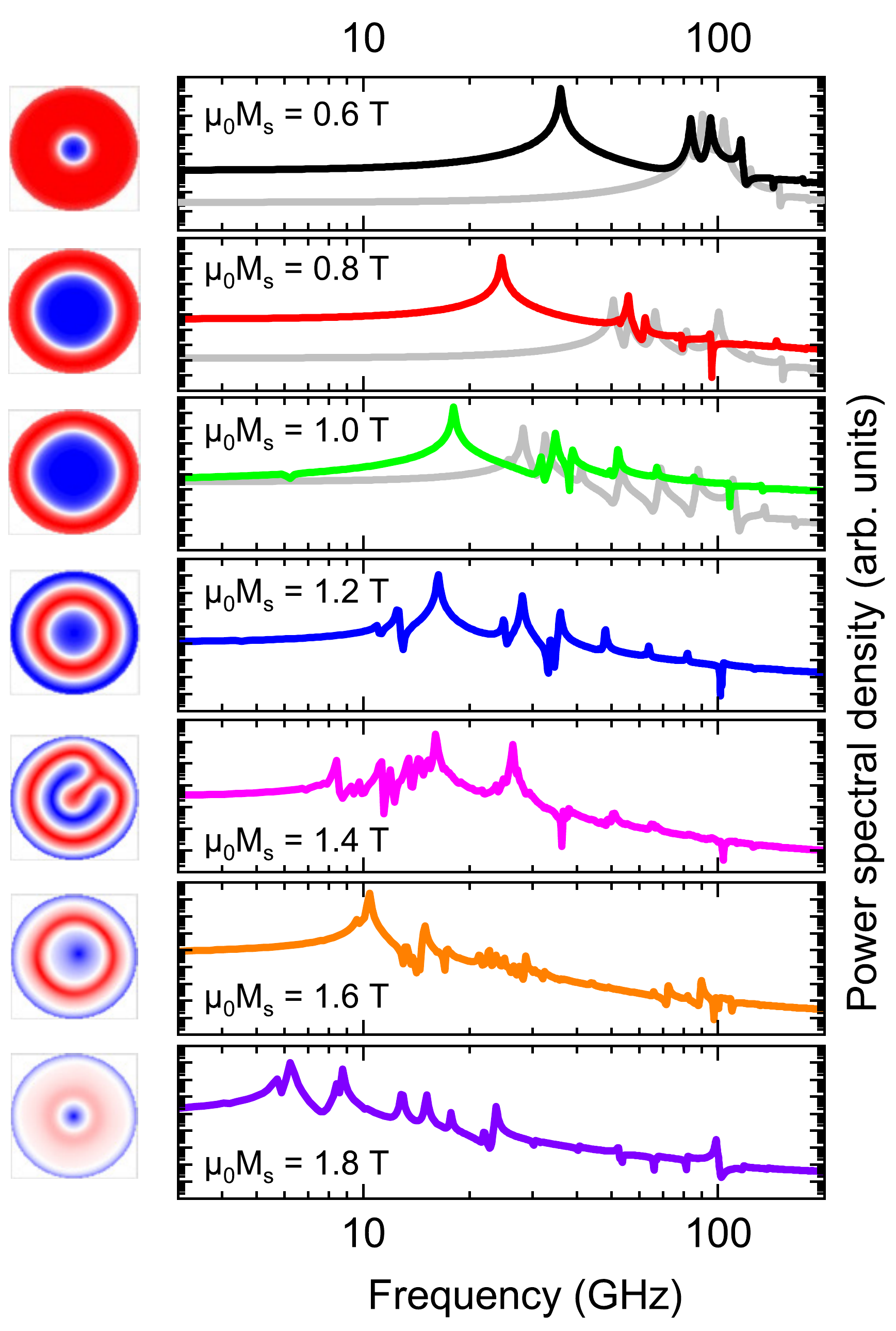}
\caption{Power spectra for SAF nanodisk with interfacial DMI, $K_{\mathrm{u}}=1\,$MJ/m$^3$ and varying saturation magnetization $M_{\mathrm{s}}$ after excitation with ac magnetic field pulse applied along the $z$-direction perpendicular to the layer planes. Interlayer RKKY coupling strength is $\sigma=-3\times 10^{-3}\,$J/m$^2$. Gray lines represent spectra for fully magnetized states with $m_{z}=+1$ for one FM layer and $m_{z}=-1$ for the other.} 
\label{DYN_MSDEP}%
\end{figure}

In this section, we analyze the resonant eigenexcitations for selected values of the interlayer coupling strength $\sigma$ and uniaxial anisotropy constant $K_{\mathrm{u}}$, while varying the saturation magnetization $\mu_{0}M_{\mathrm{s}}$. Figure \ref{DYN_MSDEP} illustrates the resonance spectra for a SAF nanodisk with $K_{\mathrm{u}}=1\,$MJ/m$^3$ and varying saturation magnetization $\mu_{0}M_{\mathrm{s}}$. The interlayer antiferromagnetic coupling strength is $\sigma=-3\times 10^{-4}\,$J/m$^2$ and thereby the chosen states correspond to the fourth column in the phase diagram depicted in Fig.\ \ref{SAFDISK_INITCOND}(a). 

Here, the power spectral density (PSD) is shown on the $y$-axis with arbitrary units, while the $x$-axis corresponds to the frequency between $3$\,GHz and $200\,$GHz. Starting from the skyrmion state in the topmost diagram ($\mu_{0}M_{\mathrm{s}}=0.6$\,T), we observe a typical behavior for magnetic excitation spectra of SAF skyrmions in a nanostructure.\cite{Lonsky2020} Due to no external dc magnetic field being applied, we only see the anti-phase breathing mode resonance peak, here at approximately $37$\,GHz, while the in-phase breathing mode that occurs at a lower frequency is being suppressed. Thereby, the complexity of the obtained power spectra is reduced significantly. The peaks at higher frequencies correspond to hybridized modes, that is, a combination of the breathing skyrmion resonance and the radial spin-wave modes of the nanodisk.\cite{Lonsky2020, Kim2014} The gray curve has been obtained by modeling the resonant dynamics of the pure antiferromagnet state (no skyrmion present) of the nanodisk with the exact material parameters, i.e., as it is shown in Fig.\ \ref{SAFDISK_INITCOND}(b) in the top right corner. This corroborates that the higher-order peaks in the black curve are related to the spin-wave modes of the nanodisk, while the peak at $37$\,GHz is solely because of the internal eigenmode of the skyrmion (anti-phase breathing mode). 

Upon increasing the saturation magnetization, the resonance peaks shift toward lower frequencies, which is also in agreement with our previous work.\cite{Lonsky2020} When the saturation magnetization is high enough, such that a skyrmionium state occurs, additional peaks appear in the spectrum. This has been also observed for skyrmioniums in FM nanostructures.\cite{Song2019, Ponsudana_2022} The complexity of the resonance spectrum increases even further when considering a C-shaped skyrmion, as for the case of $\mu_{0}M_{\mathrm{s}}=1.4$\,T. Due to the lack of radial symmetry, the magnetic excitation spectrum exhibits many more resonances. By contrast, the two highest saturation magnetization values lead to the formation of another type of incomplete skyrmionium (in analogy to incomplete skyrmions, cf.\ Refs.\ \cite{Winkler_2021, Beg_2015}) with relatively strong in-plane magnetization and topological charge that is not that of a skyrmionium, which in turn has fewer peaks than the non-circular, C-shaped skyrmion. However, as expected, the dynamics becomes even slower due to the increase in saturation magnetization. We conclude that the dynamic fingerprint of distinct magnetic configurations is unique and could be used for the experimental detection of magnetic states in SAFs. In detail, broadband ferromagnetic resonance measurements in combination with micromagnetic modeling would allow to precisely determine parameters such as the interlayer coupling strength $\sigma$ or the exact magnetic state of the sample. As will be discussed in the final section of this paper, such measurements are even conceivable for clusters of skyrmions in SAF nanostructures. 

\subsection{Magnetic Excitation Spectrum of Skyrmion Clusters in a Square-Shaped SAF Geometry}
\begin{figure}
\centering
\includegraphics[width=8.5 cm]{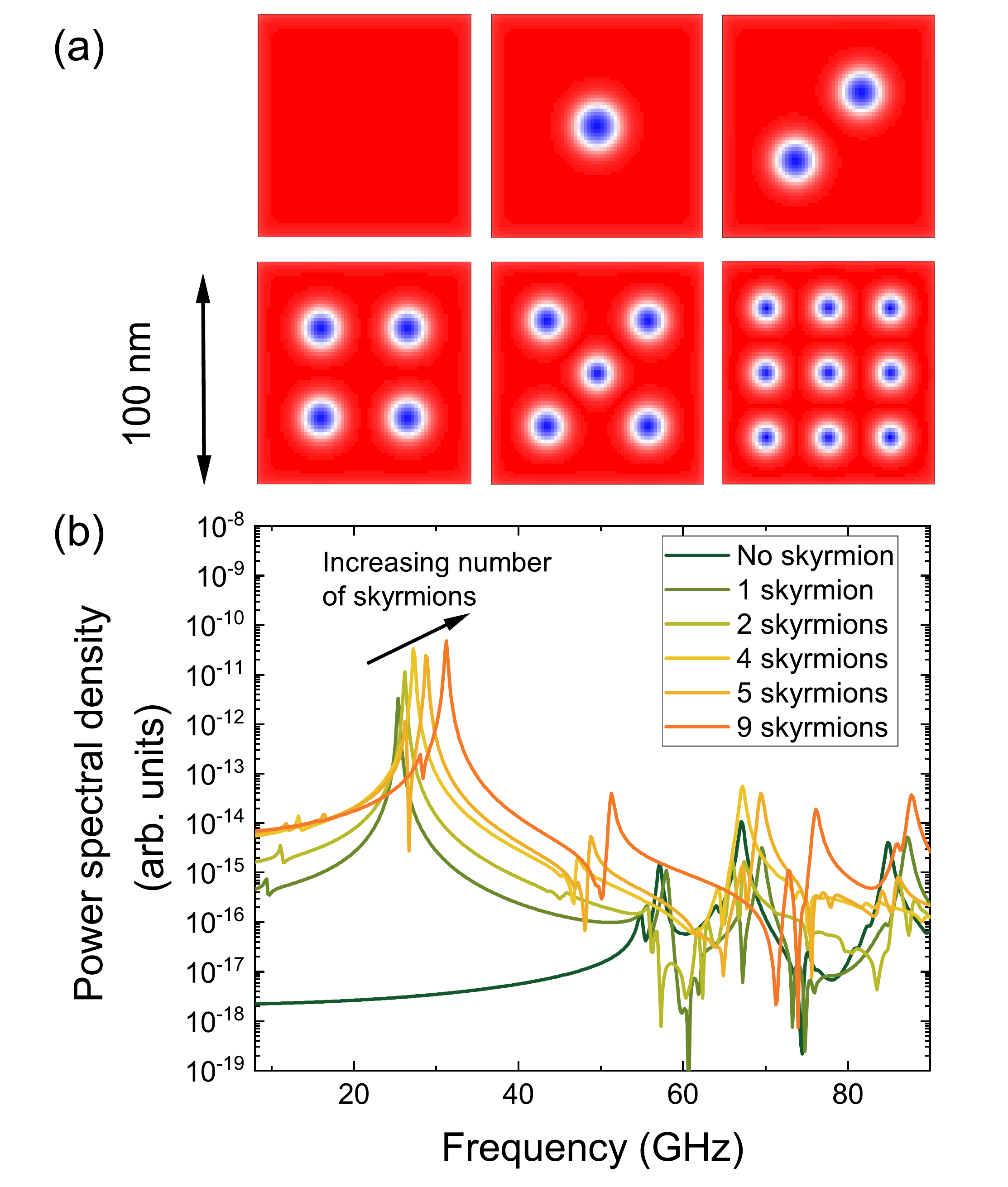}
\caption{(a) Various magnetic configurations of a SAF trilayer with square-shaped geometry, lateral size: $100\,\mathrm{nm} \times 100\,$nm. Number of skyrmions in the cluster varies from 0 to 9 and was controlled by defining appropriate initial magnetization states prior to energy minimization. Depicted is only the top layer of the SAF. (b) Magnetic excitation spectra for the six different configurations, where the resonance frequency clearly increases with the number of skyrmions.} 
\label{DYN_CLUSTER}%
\end{figure}

In the last part of this paper, we present the dynamic fingerprint of skyrmion clusters in a square-shaped SAF nanostructure with interfacial DMI. In analogy to previous works for FM nanostructures,\cite{Tejo_2020, Tejo_2020a, Saavedra_2021} it can be shown that the resonant dynamics depends on the skyrmion number. Figure \ref{DYN_CLUSTER}(a) illustrates six different relaxed magnetic states that have been obtained by using different initial magnetization conditions in order to achieve different sizes of skyrmion clusters in the square-shaped geometry. Clearly, during the energy minimization process the individual skyrmions are being arranged such that the distance between two adjacent spin textures is maximized while the arrangement should be commensurate with the geometric constriction of the nanostructure.  

In Fig.\ \ref{DYN_CLUSTER}(b) it can be observed that the resonance peaks related to the anti-phase skyrmion breathing modes shift toward higher frequencies for the case of increasing numbers of skyrmions in the cluster. As we mentioned for the case of a single skyrmion further above, the application of a symmetry breaking dc magnetic field would lead to the existence of both in-phase and anti-phase peaks, as observed in Ref.\ \cite{Lonsky2020}. In analogy to previous studies by Tejo \textit{et al.}, additional peaks in the excitation spectra for larger numbers of skyrmions are related to nonradial breathing modes due to the interaction of skyrmions in the confined geometry.\cite{Tejo_2020}
To summarize, by applying broadband ferromagnetic resonance measurements to SAF nanostructures that host magnetic skyrmions, one is able to sense and determine the number of these magnetic whirls. Furthermore, the results of this work and Ref.\ \cite{Lonsky2020} imply that varying both the number of skyrmions in a SAF nanostructure and the interlayer antiferromagnetic coupling strength allow for a highly tunable resonant dynamics which could have further applications in the field of magnonics. In this context, especially the tunability of the resonance peaks represents a major advantage over conventional bulk antiferromagnets.   

\section{Summary and Conclusion}
In this work, we have numerically studied static and dynamic properties of SAF nanostructures with interfacial DMI. It was demonstrated that there are both commonalities and significant differences in the phase diagrams of these systems in comparison to FM nanostructures without magnetic compensation. Major differences include the existence of C-shaped skyrmions as well as multi-domain states and the absence of magnetic vortex states in the investigated SAF nanodisks, where magnetic radial vortices are predominant. Furthermore, we have discussed the role of the interlayer antiferromagnetic RKKY interaction for the static properties by considering SAF nanodisks with both strong and weak coupling strength. Beyond that, we found that each magnetic configuration has its characteristic dynamic fingerprint when exposing the SAF to a microwave out-of-plane magnetic field pulse and determining the corresponding resonance modes. In particular, it has been demonstrated that even the number of skyrmions in such a SAF nanostructure leads to systematically varying dynamic properties, similar to the cases of ferromagnetic nanostructures\cite{Tejo_2020, Tejo_2020a, Saavedra_2021} and topological defects in artificial spin ice.\cite{Gliga2013} Therefore, we propose that broadband ferromagnetic resonance measurements of such systems may serve as a suitable tool to detect the precise magnetic configuration, which is typically more challenging in SAFs due to the magnetic compensation and the concomitant absence of a net magnetization. Furthermore, it was proposed that internal resonance modes of skyrmions in antiferromagnetic materials may be observed in (micro-focused) Brillouin light scattering experiments.\cite{Kravchuk2019, Lonsky2020a}
In the future, atomistic spin model simulations may be used for more realistic systems which include interface roughness, intermixing of adjacent layers, or other types of material defects.\cite{Evans2014} Thus far, we have only investigated idealized nanostructure geometries without any inhomogeneities.        

\section*{Acknowledgements}
M.\ L.\ acknowledges the financial support by the German Science Foundation (Deutsche Forschungsgemeinschaft, DFG) through the research fellowship LO 2584/1-1. This research was partially supported by the NSF through the University of Illinois at Urbana-Champaign Materials Research Science and Engineering Center DMR-1720633 and was carried out in part in the Materials Research Laboratory Central Research Facilities, University of Illinois.

\bibliography{Skyrmion_literature}

\begin{thebibliography}{44}%
\makeatletter
\providecommand \@ifxundefined [1]{%
 \@ifx{#1\undefined}
}%
\providecommand \@ifnum [1]{%
 \ifnum #1\expandafter \@firstoftwo
 \else \expandafter \@secondoftwo
 \fi
}%
\providecommand \@ifx [1]{%
 \ifx #1\expandafter \@firstoftwo
 \else \expandafter \@secondoftwo
 \fi
}%
\providecommand \natexlab [1]{#1}%
\providecommand \enquote  [1]{``#1''}%
\providecommand \bibnamefont  [1]{#1}%
\providecommand \bibfnamefont [1]{#1}%
\providecommand \citenamefont [1]{#1}%
\providecommand \href@noop [0]{\@secondoftwo}%
\providecommand \href [0]{\begingroup \@sanitize@url \@href}%
\providecommand \@href[1]{\@@startlink{#1}\@@href}%
\providecommand \@@href[1]{\endgroup#1\@@endlink}%
\providecommand \@sanitize@url [0]{\catcode `\\12\catcode `\$12\catcode
  `\&12\catcode `\#12\catcode `\^12\catcode `\_12\catcode `\%12\relax}%
\providecommand \@@startlink[1]{}%
\providecommand \@@endlink[0]{}%
\providecommand \url  [0]{\begingroup\@sanitize@url \@url }%
\providecommand \@url [1]{\endgroup\@href {#1}{\urlprefix }}%
\providecommand \urlprefix  [0]{URL }%
\providecommand \Eprint [0]{\href }%
\providecommand \doibase [0]{https://doi.org/}%
\providecommand \selectlanguage [0]{\@gobble}%
\providecommand \bibinfo  [0]{\@secondoftwo}%
\providecommand \bibfield  [0]{\@secondoftwo}%
\providecommand \translation [1]{[#1]}%
\providecommand \BibitemOpen [0]{}%
\providecommand \bibitemStop [0]{}%
\providecommand \bibitemNoStop [0]{.\EOS\space}%
\providecommand \EOS [0]{\spacefactor3000\relax}%
\providecommand \BibitemShut  [1]{\csname bibitem#1\endcsname}%
\let\auto@bib@innerbib\@empty
\bibitem [{\citenamefont {Duine}\ \emph {et~al.}(2018)\citenamefont {Duine},
  \citenamefont {Lee}, \citenamefont {Parkin},\ and\ \citenamefont
  {Stiles}}]{Duine2018}%
  \BibitemOpen
  \bibfield  {author} {\bibinfo {author} {\bibfnamefont {R.~A.}\ \bibnamefont
  {Duine}}, \bibinfo {author} {\bibfnamefont {K.-J.}\ \bibnamefont {Lee}},
  \bibinfo {author} {\bibfnamefont {S.~S.~P.}\ \bibnamefont {Parkin}},\ and\
  \bibinfo {author} {\bibfnamefont {M.~D.}\ \bibnamefont {Stiles}},\ }\bibfield
   {title} {\enquote {\bibinfo {title} {Synthetic antiferromagnetic
  spintronics},}\ }\href {https://doi.org/10.1038/s41567-018-0050-y} {\bibfield
   {journal} {\bibinfo  {journal} {Nature Physics}\ }\textbf {\bibinfo {volume}
  {14}},\ \bibinfo {pages} {217--219} (\bibinfo {year} {2018})}\BibitemShut
  {NoStop}%
\bibitem [{\citenamefont {Waring}\ \emph {et~al.}(2020)\citenamefont {Waring},
  \citenamefont {Johansson}, \citenamefont {Vera-Marun},\ and\ \citenamefont
  {Thomson}}]{Waring2020}%
  \BibitemOpen
  \bibfield  {author} {\bibinfo {author} {\bibfnamefont {H.~J.}\ \bibnamefont
  {Waring}}, \bibinfo {author} {\bibfnamefont {N.~A.~B.}\ \bibnamefont
  {Johansson}}, \bibinfo {author} {\bibfnamefont {I.~J.}\ \bibnamefont
  {Vera-Marun}},\ and\ \bibinfo {author} {\bibfnamefont {T.}~\bibnamefont
  {Thomson}},\ }\bibfield  {title} {\enquote {\bibinfo {title} {Zero-field
  optic mode beyond 20~{GHz} in a synthetic antiferromagnet},}\ }\href
  {https://doi.org/10.1103/physrevapplied.13.034035} {\bibfield  {journal}
  {\bibinfo  {journal} {Physical Review Applied}\ }\textbf {\bibinfo {volume}
  {13}},\ \bibinfo {pages} {034035} (\bibinfo {year} {2020})}\BibitemShut
  {NoStop}%
\bibitem [{\citenamefont {Dohi}\ \emph {et~al.}(2019)\citenamefont {Dohi},
  \citenamefont {DuttaGupta}, \citenamefont {Fukami},\ and\ \citenamefont
  {Ohno}}]{Dohi2019}%
  \BibitemOpen
  \bibfield  {author} {\bibinfo {author} {\bibfnamefont {T.}~\bibnamefont
  {Dohi}}, \bibinfo {author} {\bibfnamefont {S.}~\bibnamefont {DuttaGupta}},
  \bibinfo {author} {\bibfnamefont {S.}~\bibnamefont {Fukami}},\ and\ \bibinfo
  {author} {\bibfnamefont {H.}~\bibnamefont {Ohno}},\ }\bibfield  {title}
  {\enquote {\bibinfo {title} {Formation and current-induced motion of
  synthetic antiferromagnetic skyrmion bubbles},}\ }\href
  {https://doi.org/10.1038/s41467-019-13182-6} {\bibfield  {journal} {\bibinfo
  {journal} {Nature Communications}\ }\textbf {\bibinfo {volume} {10}},\
  \bibinfo {pages} {5153} (\bibinfo {year} {2019})}\BibitemShut {NoStop}%
\bibitem [{\citenamefont {Dai}\ and\ \citenamefont {Ma}(2021)}]{Dai2021}%
  \BibitemOpen
  \bibfield  {author} {\bibinfo {author} {\bibfnamefont {C.}~\bibnamefont
  {Dai}}\ and\ \bibinfo {author} {\bibfnamefont {F.}~\bibnamefont {Ma}},\
  }\bibfield  {title} {\enquote {\bibinfo {title} {Strong
  magnon{\textendash}magnon coupling in synthetic antiferromagnets},}\ }\href
  {https://doi.org/10.1063/5.0041431} {\bibfield  {journal} {\bibinfo
  {journal} {Applied Physics Letters}\ }\textbf {\bibinfo {volume} {118}},\
  \bibinfo {pages} {112405} (\bibinfo {year} {2021})}\BibitemShut {NoStop}%
\bibitem [{\citenamefont {Sorokin}\ \emph {et~al.}(2020)\citenamefont
  {Sorokin}, \citenamefont {Gallardo}, \citenamefont {Fowley}, \citenamefont
  {Lenz}, \citenamefont {Titova}, \citenamefont {Atcheson}, \citenamefont
  {Dennehy}, \citenamefont {Rode}, \citenamefont {Fassbender}, \citenamefont
  {Lindner},\ and\ \citenamefont {Deac}}]{Sorokin2020}%
  \BibitemOpen
  \bibfield  {author} {\bibinfo {author} {\bibfnamefont {S.}~\bibnamefont
  {Sorokin}}, \bibinfo {author} {\bibfnamefont {R.~A.}\ \bibnamefont
  {Gallardo}}, \bibinfo {author} {\bibfnamefont {C.}~\bibnamefont {Fowley}},
  \bibinfo {author} {\bibfnamefont {K.}~\bibnamefont {Lenz}}, \bibinfo {author}
  {\bibfnamefont {A.}~\bibnamefont {Titova}}, \bibinfo {author} {\bibfnamefont
  {G.~Y.~P.}\ \bibnamefont {Atcheson}}, \bibinfo {author} {\bibfnamefont
  {G.}~\bibnamefont {Dennehy}}, \bibinfo {author} {\bibfnamefont
  {K.}~\bibnamefont {Rode}}, \bibinfo {author} {\bibfnamefont {J.}~\bibnamefont
  {Fassbender}}, \bibinfo {author} {\bibfnamefont {J.}~\bibnamefont
  {Lindner}},\ and\ \bibinfo {author} {\bibfnamefont {A.~M.}\ \bibnamefont
  {Deac}},\ }\bibfield  {title} {\enquote {\bibinfo {title} {Magnetization
  dynamics in synthetic antiferromagnets: Role of dynamical energy and mutual
  spin pumping},}\ }\href {https://doi.org/10.1103/physrevb.101.144410}
  {\bibfield  {journal} {\bibinfo  {journal} {Physical Review B}\ }\textbf
  {\bibinfo {volume} {101}},\ \bibinfo {pages} {144410} (\bibinfo {year}
  {2020})}\BibitemShut {NoStop}%
\bibitem [{\citenamefont {Parkin}, \citenamefont {Bhadra},\ and\ \citenamefont
  {Roche}(1991)}]{Parkin1991}%
  \BibitemOpen
  \bibfield  {author} {\bibinfo {author} {\bibfnamefont {S.~S.~P.}\
  \bibnamefont {Parkin}}, \bibinfo {author} {\bibfnamefont {R.}~\bibnamefont
  {Bhadra}},\ and\ \bibinfo {author} {\bibfnamefont {K.~P.}\ \bibnamefont
  {Roche}},\ }\bibfield  {title} {\enquote {\bibinfo {title} {Oscillatory
  magnetic exchange coupling through thin copper layers},}\ }\href
  {https://doi.org/10.1103/PhysRevLett.66.2152} {\bibfield  {journal} {\bibinfo
   {journal} {Physical Review Letters}\ }\textbf {\bibinfo {volume} {66}},\
  \bibinfo {pages} {2152--2155} (\bibinfo {year} {1991})}\BibitemShut {NoStop}%
\bibitem [{\citenamefont {Bruno}\ and\ \citenamefont
  {Chappert}(1991)}]{Bruno1991}%
  \BibitemOpen
  \bibfield  {author} {\bibinfo {author} {\bibfnamefont {P.}~\bibnamefont
  {Bruno}}\ and\ \bibinfo {author} {\bibfnamefont {C.}~\bibnamefont
  {Chappert}},\ }\bibfield  {title} {\enquote {\bibinfo {title} {Oscillatory
  coupling between ferromagnetic layers separated by a nonmagnetic metal
  spacer},}\ }\href {https://doi.org/10.1103/PhysRevLett.67.1602} {\bibfield
  {journal} {\bibinfo  {journal} {Physical Review Letters}\ }\textbf {\bibinfo
  {volume} {67}},\ \bibinfo {pages} {1602--1605} (\bibinfo {year}
  {1991})}\BibitemShut {NoStop}%
\bibitem [{\citenamefont {Dzyaloshinskii}(1958)}]{Dzyaloshinskii1958}%
  \BibitemOpen
  \bibfield  {author} {\bibinfo {author} {\bibfnamefont {I.}~\bibnamefont
  {Dzyaloshinskii}},\ }\bibfield  {title} {\enquote {\bibinfo {title} {A
  thermodynamic theory of {\textquotedblleft}weak{\textquotedblright}
  ferromagnetism of antiferromagnetics},}\ }\href
  {https://doi.org/10.1016/0022-3697(58)90076-3} {\bibfield  {journal}
  {\bibinfo  {journal} {Journal of Physics and Chemistry of Solids}\ }\textbf
  {\bibinfo {volume} {4}},\ \bibinfo {pages} {241--255} (\bibinfo {year}
  {1958})}\BibitemShut {NoStop}%
\bibitem [{\citenamefont {Moriya}(1960)}]{Moriya1960}%
  \BibitemOpen
  \bibfield  {author} {\bibinfo {author} {\bibfnamefont {T.}~\bibnamefont
  {Moriya}},\ }\bibfield  {title} {\enquote {\bibinfo {title} {Anisotropic
  superexchange interaction and weak ferromagnetism},}\ }\href
  {https://doi.org/10.1103/PhysRev.120.91} {\bibfield  {journal} {\bibinfo
  {journal} {Physical Review}\ }\textbf {\bibinfo {volume} {120}},\ \bibinfo
  {pages} {91--98} (\bibinfo {year} {1960})}\BibitemShut {NoStop}%
\bibitem [{\citenamefont {Legrand}\ \emph {et~al.}(2020)\citenamefont
  {Legrand}, \citenamefont {Maccariello}, \citenamefont {Ajejas}, \citenamefont
  {Collin}, \citenamefont {Vecchiola}, \citenamefont {Bouzehouane},
  \citenamefont {Reyren}, \citenamefont {Cros},\ and\ \citenamefont
  {Fert}}]{Legrand2020}%
  \BibitemOpen
  \bibfield  {author} {\bibinfo {author} {\bibfnamefont {W.}~\bibnamefont
  {Legrand}}, \bibinfo {author} {\bibfnamefont {D.}~\bibnamefont
  {Maccariello}}, \bibinfo {author} {\bibfnamefont {F.}~\bibnamefont {Ajejas}},
  \bibinfo {author} {\bibfnamefont {S.}~\bibnamefont {Collin}}, \bibinfo
  {author} {\bibfnamefont {A.}~\bibnamefont {Vecchiola}}, \bibinfo {author}
  {\bibfnamefont {K.}~\bibnamefont {Bouzehouane}}, \bibinfo {author}
  {\bibfnamefont {N.}~\bibnamefont {Reyren}}, \bibinfo {author} {\bibfnamefont
  {V.}~\bibnamefont {Cros}},\ and\ \bibinfo {author} {\bibfnamefont
  {A.}~\bibnamefont {Fert}},\ }\bibfield  {title} {\enquote {\bibinfo {title}
  {Room-temperature stabilization of antiferromagnetic skyrmions in synthetic
  antiferromagnets},}\ }\href {https://doi.org/10.1038/s41563-019-0468-3}
  {\bibfield  {journal} {\bibinfo  {journal} {Nature Materials}\ }\textbf
  {\bibinfo {volume} {19}},\ \bibinfo {pages} {34--42} (\bibinfo {year}
  {2020})}\BibitemShut {NoStop}%
\bibitem [{\citenamefont {Chen}\ \emph {et~al.}(2020)\citenamefont {Chen},
  \citenamefont {Gao}, \citenamefont {Zhang}, \citenamefont {Zhang},
  \citenamefont {Yin}, \citenamefont {Chen}, \citenamefont {Zhou},
  \citenamefont {Zhou}, \citenamefont {Xia}, \citenamefont {Zhou},
  \citenamefont {Wang}, \citenamefont {Pan}, \citenamefont {Zhang},\ and\
  \citenamefont {Song}}]{Chen2020}%
  \BibitemOpen
  \bibfield  {author} {\bibinfo {author} {\bibfnamefont {R.}~\bibnamefont
  {Chen}}, \bibinfo {author} {\bibfnamefont {Y.}~\bibnamefont {Gao}}, \bibinfo
  {author} {\bibfnamefont {X.}~\bibnamefont {Zhang}}, \bibinfo {author}
  {\bibfnamefont {R.}~\bibnamefont {Zhang}}, \bibinfo {author} {\bibfnamefont
  {S.}~\bibnamefont {Yin}}, \bibinfo {author} {\bibfnamefont {X.}~\bibnamefont
  {Chen}}, \bibinfo {author} {\bibfnamefont {X.}~\bibnamefont {Zhou}}, \bibinfo
  {author} {\bibfnamefont {Y.}~\bibnamefont {Zhou}}, \bibinfo {author}
  {\bibfnamefont {J.}~\bibnamefont {Xia}}, \bibinfo {author} {\bibfnamefont
  {Y.}~\bibnamefont {Zhou}}, \bibinfo {author} {\bibfnamefont {S.}~\bibnamefont
  {Wang}}, \bibinfo {author} {\bibfnamefont {F.}~\bibnamefont {Pan}}, \bibinfo
  {author} {\bibfnamefont {Y.}~\bibnamefont {Zhang}},\ and\ \bibinfo {author}
  {\bibfnamefont {C.}~\bibnamefont {Song}},\ }\bibfield  {title} {\enquote
  {\bibinfo {title} {Realization of isolated and high-density skyrmions at room
  temperature in uncompensated synthetic antiferromagnets},}\ }\href
  {https://doi.org/10.1021/acs.nanolett.0c00116} {\bibfield  {journal}
  {\bibinfo  {journal} {Nano Letters}\ }\textbf {\bibinfo {volume} {20}},\
  \bibinfo {pages} {3299--3305} (\bibinfo {year} {2020})}\BibitemShut {NoStop}%
\bibitem [{\citenamefont {Juge}\ \emph {et~al.}(2021)\citenamefont {Juge},
  \citenamefont {Sisodia}, \citenamefont {Larrañaga}, \citenamefont {Zhang},
  \citenamefont {Pham}, \citenamefont {Rana}, \citenamefont {Sarpi},
  \citenamefont {Mille}, \citenamefont {Stanescu}, \citenamefont {Belkhou},
  \citenamefont {Mawass}, \citenamefont {Novakovic-Marinkovic}, \citenamefont
  {Kronast}, \citenamefont {Weigand}, \citenamefont {Gräfe}, \citenamefont
  {Wintz}, \citenamefont {Finizio}, \citenamefont {Raabe}, \citenamefont
  {Aballe}, \citenamefont {Foerster}, \citenamefont {Belmeguenai},
  \citenamefont {Buda-Prejbeanu}, \citenamefont {Shaw}, \citenamefont
  {Nembach}, \citenamefont {Ranno}, \citenamefont {Gaudin},\ and\ \citenamefont
  {Boulle}}]{Juge_2021}%
  \BibitemOpen
  \bibfield  {author} {\bibinfo {author} {\bibfnamefont {R.}~\bibnamefont
  {Juge}}, \bibinfo {author} {\bibfnamefont {N.}~\bibnamefont {Sisodia}},
  \bibinfo {author} {\bibfnamefont {J.~U.}\ \bibnamefont {Larrañaga}},
  \bibinfo {author} {\bibfnamefont {Q.}~\bibnamefont {Zhang}}, \bibinfo
  {author} {\bibfnamefont {V.~T.}\ \bibnamefont {Pham}}, \bibinfo {author}
  {\bibfnamefont {K.~G.}\ \bibnamefont {Rana}}, \bibinfo {author}
  {\bibfnamefont {B.}~\bibnamefont {Sarpi}}, \bibinfo {author} {\bibfnamefont
  {N.}~\bibnamefont {Mille}}, \bibinfo {author} {\bibfnamefont
  {S.}~\bibnamefont {Stanescu}}, \bibinfo {author} {\bibfnamefont
  {R.}~\bibnamefont {Belkhou}}, \bibinfo {author} {\bibfnamefont {M.-A.}\
  \bibnamefont {Mawass}}, \bibinfo {author} {\bibfnamefont {N.}~\bibnamefont
  {Novakovic-Marinkovic}}, \bibinfo {author} {\bibfnamefont {F.}~\bibnamefont
  {Kronast}}, \bibinfo {author} {\bibfnamefont {M.}~\bibnamefont {Weigand}},
  \bibinfo {author} {\bibfnamefont {J.}~\bibnamefont {Gräfe}}, \bibinfo
  {author} {\bibfnamefont {S.}~\bibnamefont {Wintz}}, \bibinfo {author}
  {\bibfnamefont {S.}~\bibnamefont {Finizio}}, \bibinfo {author} {\bibfnamefont
  {J.}~\bibnamefont {Raabe}}, \bibinfo {author} {\bibfnamefont
  {L.}~\bibnamefont {Aballe}}, \bibinfo {author} {\bibfnamefont
  {M.}~\bibnamefont {Foerster}}, \bibinfo {author} {\bibfnamefont
  {M.}~\bibnamefont {Belmeguenai}}, \bibinfo {author} {\bibfnamefont
  {L.}~\bibnamefont {Buda-Prejbeanu}}, \bibinfo {author} {\bibfnamefont
  {J.~M.}\ \bibnamefont {Shaw}}, \bibinfo {author} {\bibfnamefont {H.~T.}\
  \bibnamefont {Nembach}}, \bibinfo {author} {\bibfnamefont {L.}~\bibnamefont
  {Ranno}}, \bibinfo {author} {\bibfnamefont {G.}~\bibnamefont {Gaudin}},\ and\
  \bibinfo {author} {\bibfnamefont {O.}~\bibnamefont {Boulle}},\ }\bibfield
  {title} {\enquote {\bibinfo {title} {Skyrmions in synthetic antiferromagnets
  and their nucleation via electrical current and ultrafast laser
  illumination},}\ }\href@noop {} {\  (\bibinfo {year} {2021})},\ \Eprint
  {https://arxiv.org/abs/https://arxiv.org/abs/2111.11878}
  {arXiv:https://arxiv.org/abs/2111.11878 [cond-mat.mtrl-sci]} \BibitemShut
  {NoStop}%
\bibitem [{\citenamefont {Mathur}, \citenamefont {Stolt},\ and\ \citenamefont
  {Jin}(2019)}]{Mathur2019}%
  \BibitemOpen
  \bibfield  {author} {\bibinfo {author} {\bibfnamefont {N.}~\bibnamefont
  {Mathur}}, \bibinfo {author} {\bibfnamefont {M.~J.}\ \bibnamefont {Stolt}},\
  and\ \bibinfo {author} {\bibfnamefont {S.}~\bibnamefont {Jin}},\ }\bibfield
  {title} {\enquote {\bibinfo {title} {Magnetic skyrmions in nanostructures of
  non-centrosymmetric materials},}\ }\href {https://doi.org/10.1063/1.5130423}
  {\bibfield  {journal} {\bibinfo  {journal} {{APL} Materials}\ }\textbf
  {\bibinfo {volume} {7}},\ \bibinfo {pages} {120703} (\bibinfo {year}
  {2019})}\BibitemShut {NoStop}%
\bibitem [{\citenamefont {Lonsky}\ and\ \citenamefont
  {Hoffmann}(2020{\natexlab{a}})}]{Lonsky2020}%
  \BibitemOpen
  \bibfield  {author} {\bibinfo {author} {\bibfnamefont {M.}~\bibnamefont
  {Lonsky}}\ and\ \bibinfo {author} {\bibfnamefont {A.}~\bibnamefont
  {Hoffmann}},\ }\bibfield  {title} {\enquote {\bibinfo {title} {Coupled
  skyrmion breathing modes in synthetic ferri- and antiferromagnets},}\ }\href
  {https://doi.org/10.1103/physrevb.102.104403} {\bibfield  {journal} {\bibinfo
   {journal} {Physical Review B}\ }\textbf {\bibinfo {volume} {102}},\ \bibinfo
  {pages} {104403} (\bibinfo {year} {2020}{\natexlab{a}})}\BibitemShut
  {NoStop}%
\bibitem [{\citenamefont {Lonsky}\ and\ \citenamefont
  {Hoffmann}(2020{\natexlab{b}})}]{Lonsky2020a}%
  \BibitemOpen
  \bibfield  {author} {\bibinfo {author} {\bibfnamefont {M.}~\bibnamefont
  {Lonsky}}\ and\ \bibinfo {author} {\bibfnamefont {A.}~\bibnamefont
  {Hoffmann}},\ }\bibfield  {title} {\enquote {\bibinfo {title} {Dynamic
  excitations of chiral magnetic textures},}\ }\href
  {https://doi.org/10.1063/5.0027042} {\bibfield  {journal} {\bibinfo
  {journal} {{APL} Materials}\ }\textbf {\bibinfo {volume} {8}},\ \bibinfo
  {pages} {100903} (\bibinfo {year} {2020}{\natexlab{b}})}\BibitemShut
  {NoStop}%
\bibitem [{\citenamefont {Qiu}\ \emph {et~al.}(2021)\citenamefont {Qiu},
  \citenamefont {Shen}, \citenamefont {Zhang}, \citenamefont {Zhou},
  \citenamefont {Zhao}, \citenamefont {Xia}, \citenamefont {Luo},\ and\
  \citenamefont {Liu}}]{Qiu2021}%
  \BibitemOpen
  \bibfield  {author} {\bibinfo {author} {\bibfnamefont {L.}~\bibnamefont
  {Qiu}}, \bibinfo {author} {\bibfnamefont {L.}~\bibnamefont {Shen}}, \bibinfo
  {author} {\bibfnamefont {X.}~\bibnamefont {Zhang}}, \bibinfo {author}
  {\bibfnamefont {Y.}~\bibnamefont {Zhou}}, \bibinfo {author} {\bibfnamefont
  {G.}~\bibnamefont {Zhao}}, \bibinfo {author} {\bibfnamefont {W.}~\bibnamefont
  {Xia}}, \bibinfo {author} {\bibfnamefont {H.-B.}\ \bibnamefont {Luo}},\ and\
  \bibinfo {author} {\bibfnamefont {J.~P.}\ \bibnamefont {Liu}},\ }\bibfield
  {title} {\enquote {\bibinfo {title} {Interlayer coupling effect on skyrmion
  dynamics in synthetic antiferromagnets},}\ }\href
  {https://doi.org/10.1063/5.0039470} {\bibfield  {journal} {\bibinfo
  {journal} {Applied Physics Letters}\ }\textbf {\bibinfo {volume} {118}},\
  \bibinfo {pages} {082403} (\bibinfo {year} {2021})}\BibitemShut {NoStop}%
\bibitem [{\citenamefont {Winkler}\ \emph {et~al.}(2021)\citenamefont
  {Winkler}, \citenamefont {Litzius}, \citenamefont {de~Lucia}, \citenamefont
  {Wei{\ss}enhofer}, \citenamefont {Fangohr},\ and\ \citenamefont
  {Kläui}}]{Winkler_2021}%
  \BibitemOpen
  \bibfield  {author} {\bibinfo {author} {\bibfnamefont {T.~B.}\ \bibnamefont
  {Winkler}}, \bibinfo {author} {\bibfnamefont {K.}~\bibnamefont {Litzius}},
  \bibinfo {author} {\bibfnamefont {A.}~\bibnamefont {de~Lucia}}, \bibinfo
  {author} {\bibfnamefont {M.}~\bibnamefont {Wei{\ss}enhofer}}, \bibinfo
  {author} {\bibfnamefont {H.}~\bibnamefont {Fangohr}},\ and\ \bibinfo {author}
  {\bibfnamefont {M.}~\bibnamefont {Kläui}},\ }\bibfield  {title} {\enquote
  {\bibinfo {title} {Skyrmion states in disk geometry},}\ }\href
  {https://doi.org/10.1103/PhysRevApplied.16.044014} {\bibfield  {journal}
  {\bibinfo  {journal} {Physical Review Applied}\ }\textbf {\bibinfo {volume}
  {16}},\ \bibinfo {pages} {044014} (\bibinfo {year} {2021})}\BibitemShut
  {NoStop}%
\bibitem [{\citenamefont {Ji}\ \emph {et~al.}(2021)\citenamefont {Ji},
  \citenamefont {Zhao}, \citenamefont {Hu}, \citenamefont {Chen}, \citenamefont
  {Chen},\ and\ \citenamefont {Zhang}}]{Ji2021}%
  \BibitemOpen
  \bibfield  {author} {\bibinfo {author} {\bibfnamefont {L.}~\bibnamefont
  {Ji}}, \bibinfo {author} {\bibfnamefont {R.}~\bibnamefont {Zhao}}, \bibinfo
  {author} {\bibfnamefont {C.}~\bibnamefont {Hu}}, \bibinfo {author}
  {\bibfnamefont {W.}~\bibnamefont {Chen}}, \bibinfo {author} {\bibfnamefont
  {Y.}~\bibnamefont {Chen}},\ and\ \bibinfo {author} {\bibfnamefont
  {X.}~\bibnamefont {Zhang}},\ }\bibfield  {title} {\enquote {\bibinfo {title}
  {Transformation from antiferromagnetic target skyrmion to antiferromagnetic
  skyrmion by unzipping process through a confined nanostructure},}\ }\href
  {https://doi.org/10.1088/1361-648x/abe079} {\bibfield  {journal} {\bibinfo
  {journal} {Journal of Physics: Condensed Matter}\ }\textbf {\bibinfo {volume}
  {33}},\ \bibinfo {pages} {425801} (\bibinfo {year} {2021})}\BibitemShut
  {NoStop}%
\bibitem [{\citenamefont {Gliga}\ \emph {et~al.}(2013)\citenamefont {Gliga},
  \citenamefont {K{\'{a}}kay}, \citenamefont {Hertel},\ and\ \citenamefont
  {Heinonen}}]{Gliga2013}%
  \BibitemOpen
  \bibfield  {author} {\bibinfo {author} {\bibfnamefont {S.}~\bibnamefont
  {Gliga}}, \bibinfo {author} {\bibfnamefont {A.}~\bibnamefont {K{\'{a}}kay}},
  \bibinfo {author} {\bibfnamefont {R.}~\bibnamefont {Hertel}},\ and\ \bibinfo
  {author} {\bibfnamefont {O.~G.}\ \bibnamefont {Heinonen}},\ }\bibfield
  {title} {\enquote {\bibinfo {title} {Spectral analysis of topological defects
  in an artificial spin-ice lattice},}\ }\href
  {https://doi.org/10.1103/PhysRevLett.110.117205} {\bibfield  {journal}
  {\bibinfo  {journal} {Physical Review Letters}\ }\textbf {\bibinfo {volume}
  {110}},\ \bibinfo {pages} {117205} (\bibinfo {year} {2013})}\BibitemShut
  {NoStop}%
\bibitem [{\citenamefont {Donahue}\ and\ \citenamefont
  {Porter}(1999)}]{Donahue1999}%
  \BibitemOpen
  \bibfield  {author} {\bibinfo {author} {\bibfnamefont {M.~J.}\ \bibnamefont
  {Donahue}}\ and\ \bibinfo {author} {\bibfnamefont {D.~G.}\ \bibnamefont
  {Porter}},\ }\href {https://doi.org/10.6028/NIST.IR.6376} {\enquote {\bibinfo
  {title} {{OOMMF} {U}ser's {G}uide, {V}ersion 1.0},}\ }\bibinfo {howpublished}
  {Interagency Report NISTIR 6376, National Institute of Standards and
  Technology, Gaithersburg, MD} (\bibinfo {year} {1999})\BibitemShut {NoStop}%
\bibitem [{\citenamefont {Rohart}\ and\ \citenamefont
  {Thiaville}(2013)}]{Rohart2013}%
  \BibitemOpen
  \bibfield  {author} {\bibinfo {author} {\bibfnamefont {S.}~\bibnamefont
  {Rohart}}\ and\ \bibinfo {author} {\bibfnamefont {A.}~\bibnamefont
  {Thiaville}},\ }\bibfield  {title} {\enquote {\bibinfo {title} {Skyrmion
  confinement in ultrathin film nanostructures in the presence of
  dzyaloshinskii-moriya interaction},}\ }\href
  {https://doi.org/10.1103/PhysRevB.88.184422} {\bibfield  {journal} {\bibinfo
  {journal} {Physical Review B}\ }\textbf {\bibinfo {volume} {88}},\ \bibinfo
  {pages} {184422} (\bibinfo {year} {2013})}\BibitemShut {NoStop}%
\bibitem [{\citenamefont {Siracusano}\ \emph {et~al.}(2016)\citenamefont
  {Siracusano}, \citenamefont {Tomasello}, \citenamefont {Giordano},
  \citenamefont {Puliafito}, \citenamefont {Azzerboni}, \citenamefont {Ozatay},
  \citenamefont {Carpentieri},\ and\ \citenamefont
  {Finocchio}}]{Siracusano_2016}%
  \BibitemOpen
  \bibfield  {author} {\bibinfo {author} {\bibfnamefont {G.}~\bibnamefont
  {Siracusano}}, \bibinfo {author} {\bibfnamefont {R.}~\bibnamefont
  {Tomasello}}, \bibinfo {author} {\bibfnamefont {A.}~\bibnamefont {Giordano}},
  \bibinfo {author} {\bibfnamefont {V.}~\bibnamefont {Puliafito}}, \bibinfo
  {author} {\bibfnamefont {B.}~\bibnamefont {Azzerboni}}, \bibinfo {author}
  {\bibfnamefont {O.}~\bibnamefont {Ozatay}}, \bibinfo {author} {\bibfnamefont
  {M.}~\bibnamefont {Carpentieri}},\ and\ \bibinfo {author} {\bibfnamefont
  {G.}~\bibnamefont {Finocchio}},\ }\bibfield  {title} {\enquote {\bibinfo
  {title} {Magnetic radial vortex stabilization and efficient manipulation
  driven by the dzyaloshinskii-moriya interaction and spin-transfer torque},}\
  }\href {https://doi.org/10.1103/PhysRevLett.117.087204} {\bibfield  {journal}
  {\bibinfo  {journal} {Physical Review Letters}\ }\textbf {\bibinfo {volume}
  {117}},\ \bibinfo {pages} {087204} (\bibinfo {year} {2016})}\BibitemShut
  {NoStop}%
\bibitem [{\citenamefont {Beg}\ \emph {et~al.}(2015)\citenamefont {Beg},
  \citenamefont {Carey}, \citenamefont {Wang}, \citenamefont
  {Cort{\'{e}}s-Ortu{\~{n}}o}, \citenamefont {Vousden}, \citenamefont
  {Bisotti}, \citenamefont {Albert}, \citenamefont {Chernyshenko},
  \citenamefont {Hovorka}, \citenamefont {Stamps},\ and\ \citenamefont
  {Fangohr}}]{Beg_2015}%
  \BibitemOpen
  \bibfield  {author} {\bibinfo {author} {\bibfnamefont {M.}~\bibnamefont
  {Beg}}, \bibinfo {author} {\bibfnamefont {R.}~\bibnamefont {Carey}}, \bibinfo
  {author} {\bibfnamefont {W.}~\bibnamefont {Wang}}, \bibinfo {author}
  {\bibfnamefont {D.}~\bibnamefont {Cort{\'{e}}s-Ortu{\~{n}}o}}, \bibinfo
  {author} {\bibfnamefont {M.}~\bibnamefont {Vousden}}, \bibinfo {author}
  {\bibfnamefont {M.-A.}\ \bibnamefont {Bisotti}}, \bibinfo {author}
  {\bibfnamefont {M.}~\bibnamefont {Albert}}, \bibinfo {author} {\bibfnamefont
  {D.}~\bibnamefont {Chernyshenko}}, \bibinfo {author} {\bibfnamefont
  {O.}~\bibnamefont {Hovorka}}, \bibinfo {author} {\bibfnamefont {R.~L.}\
  \bibnamefont {Stamps}},\ and\ \bibinfo {author} {\bibfnamefont
  {H.}~\bibnamefont {Fangohr}},\ }\bibfield  {title} {\enquote {\bibinfo
  {title} {Ground state search, hysteretic behaviour and reversal mechanism of
  skyrmionic textures in confined helimagnetic nanostructures},}\ }\href
  {https://doi.org/10.1038/srep17137} {\bibfield  {journal} {\bibinfo
  {journal} {Scientific Reports}\ }\textbf {\bibinfo {volume} {5}} (\bibinfo
  {year} {2015}),\ 10.1038/srep17137}\BibitemShut {NoStop}%
\bibitem [{\citenamefont {Kim}\ \emph {et~al.}(2014)\citenamefont {Kim},
  \citenamefont {Garcia-Sanchez}, \citenamefont {Sampaio}, \citenamefont
  {Moreau-Luchaire}, \citenamefont {Cros},\ and\ \citenamefont
  {Fert}}]{Kim2014}%
  \BibitemOpen
  \bibfield  {author} {\bibinfo {author} {\bibfnamefont {J.-V.}\ \bibnamefont
  {Kim}}, \bibinfo {author} {\bibfnamefont {F.}~\bibnamefont {Garcia-Sanchez}},
  \bibinfo {author} {\bibfnamefont {J.}~\bibnamefont {Sampaio}}, \bibinfo
  {author} {\bibfnamefont {C.}~\bibnamefont {Moreau-Luchaire}}, \bibinfo
  {author} {\bibfnamefont {V.}~\bibnamefont {Cros}},\ and\ \bibinfo {author}
  {\bibfnamefont {A.}~\bibnamefont {Fert}},\ }\bibfield  {title} {\enquote
  {\bibinfo {title} {Breathing modes of confined skyrmions in ultrathin
  magnetic dots},}\ }\href {https://doi.org/10.1103/PhysRevB.90.064410}
  {\bibfield  {journal} {\bibinfo  {journal} {Physical Review B}\ }\textbf
  {\bibinfo {volume} {90}},\ \bibinfo {pages} {064410} (\bibinfo {year}
  {2014})}\BibitemShut {NoStop}%
\bibitem [{\citenamefont {Hu}, \citenamefont {Yang},\ and\ \citenamefont
  {Chen}(2020)}]{Hu2020}%
  \BibitemOpen
  \bibfield  {author} {\bibinfo {author} {\bibfnamefont {J.-M.}\ \bibnamefont
  {Hu}}, \bibinfo {author} {\bibfnamefont {T.}~\bibnamefont {Yang}},\ and\
  \bibinfo {author} {\bibfnamefont {L.-Q.}\ \bibnamefont {Chen}},\ }\bibfield
  {title} {\enquote {\bibinfo {title} {Stability and dynamics of skyrmions in
  ultrathin magnetic nanodisks under strain},}\ }\href
  {https://doi.org/10.1016/j.actamat.2019.11.005} {\bibfield  {journal}
  {\bibinfo  {journal} {Acta Materialia}\ }\textbf {\bibinfo {volume} {183}},\
  \bibinfo {pages} {145--154} (\bibinfo {year} {2020})}\BibitemShut {NoStop}%
\bibitem [{\citenamefont {Song}\ \emph {et~al.}(2019)\citenamefont {Song},
  \citenamefont {Ma}, \citenamefont {Jin}, \citenamefont {Wang}, \citenamefont
  {Xia}, \citenamefont {Wang},\ and\ \citenamefont {Liu}}]{Song2019}%
  \BibitemOpen
  \bibfield  {author} {\bibinfo {author} {\bibfnamefont {C.}~\bibnamefont
  {Song}}, \bibinfo {author} {\bibfnamefont {Y.}~\bibnamefont {Ma}}, \bibinfo
  {author} {\bibfnamefont {C.}~\bibnamefont {Jin}}, \bibinfo {author}
  {\bibfnamefont {J.}~\bibnamefont {Wang}}, \bibinfo {author} {\bibfnamefont
  {H.}~\bibnamefont {Xia}}, \bibinfo {author} {\bibfnamefont {J.}~\bibnamefont
  {Wang}},\ and\ \bibinfo {author} {\bibfnamefont {Q.}~\bibnamefont {Liu}},\
  }\bibfield  {title} {\enquote {\bibinfo {title} {Field-tuned spin excitation
  spectrum of k$\pi$ skyrmion},}\ }\href
  {https://doi.org/10.1088/1367-2630/ab348e} {\bibfield  {journal} {\bibinfo
  {journal} {New Journal of Physics}\ }\textbf {\bibinfo {volume} {21}},\
  \bibinfo {pages} {083006} (\bibinfo {year} {2019})}\BibitemShut {NoStop}%
\bibitem [{\citenamefont {Guslienko}\ \emph {et~al.}(2002)\citenamefont
  {Guslienko}, \citenamefont {Ivanov}, \citenamefont {Novosad}, \citenamefont
  {Otani}, \citenamefont {Shima},\ and\ \citenamefont
  {Fukamichi}}]{Guslienko_2002}%
  \BibitemOpen
  \bibfield  {author} {\bibinfo {author} {\bibfnamefont {K.~Y.}\ \bibnamefont
  {Guslienko}}, \bibinfo {author} {\bibfnamefont {B.~A.}\ \bibnamefont
  {Ivanov}}, \bibinfo {author} {\bibfnamefont {V.}~\bibnamefont {Novosad}},
  \bibinfo {author} {\bibfnamefont {Y.}~\bibnamefont {Otani}}, \bibinfo
  {author} {\bibfnamefont {H.}~\bibnamefont {Shima}},\ and\ \bibinfo {author}
  {\bibfnamefont {K.}~\bibnamefont {Fukamichi}},\ }\bibfield  {title} {\enquote
  {\bibinfo {title} {Eigenfrequencies of vortex state excitations in magnetic
  submicron-size disks},}\ }\href {https://doi.org/10.1063/1.1450816}
  {\bibfield  {journal} {\bibinfo  {journal} {Journal of Applied Physics}\
  }\textbf {\bibinfo {volume} {91}},\ \bibinfo {pages} {8037} (\bibinfo {year}
  {2002})}\BibitemShut {NoStop}%
\bibitem [{\citenamefont {Choe}(2004)}]{Choe2004}%
  \BibitemOpen
  \bibfield  {author} {\bibinfo {author} {\bibfnamefont {S.-B.}\ \bibnamefont
  {Choe}},\ }\bibfield  {title} {\enquote {\bibinfo {title} {Vortex core-driven
  magnetization dynamics},}\ }\href {https://doi.org/10.1126/science.1095068}
  {\bibfield  {journal} {\bibinfo  {journal} {Science}\ }\textbf {\bibinfo
  {volume} {304}},\ \bibinfo {pages} {420--422} (\bibinfo {year}
  {2004})}\BibitemShut {NoStop}%
\bibitem [{\citenamefont {Barker}\ \emph {et~al.}(2021)\citenamefont {Barker},
  \citenamefont {Haltz}, \citenamefont {Moore},\ and\ \citenamefont
  {Marrows}}]{Barker_2021}%
  \BibitemOpen
  \bibfield  {author} {\bibinfo {author} {\bibfnamefont {C.~E.~A.}\
  \bibnamefont {Barker}}, \bibinfo {author} {\bibfnamefont {E.}~\bibnamefont
  {Haltz}}, \bibinfo {author} {\bibfnamefont {T.~A.}\ \bibnamefont {Moore}},\
  and\ \bibinfo {author} {\bibfnamefont {C.~H.}\ \bibnamefont {Marrows}},\
  }\bibfield  {title} {\enquote {\bibinfo {title} {Breathing modes of skyrmion
  strings in a synthetic antiferromagnet},}\ }\href@noop {} {\  (\bibinfo
  {year} {2021})},\ \Eprint
  {https://arxiv.org/abs/https://arxiv.org/abs/2112.05481}
  {arXiv:https://arxiv.org/abs/2112.05481 [cond-mat.mes-hall]} \BibitemShut
  {NoStop}%
\bibitem [{\citenamefont {Castell-Queralt}\ \emph {et~al.}(2022)\citenamefont
  {Castell-Queralt}, \citenamefont {Gonz{\'{a}}lez-G{\'{o}}mez}, \citenamefont
  {Del-Valle},\ and\ \citenamefont {Navau}}]{Castell_Queralt_2022}%
  \BibitemOpen
  \bibfield  {author} {\bibinfo {author} {\bibfnamefont {J.}~\bibnamefont
  {Castell-Queralt}}, \bibinfo {author} {\bibfnamefont {L.}~\bibnamefont
  {Gonz{\'{a}}lez-G{\'{o}}mez}}, \bibinfo {author} {\bibfnamefont
  {N.}~\bibnamefont {Del-Valle}},\ and\ \bibinfo {author} {\bibfnamefont
  {C.}~\bibnamefont {Navau}},\ }\bibfield  {title} {\enquote {\bibinfo {title}
  {Exploiting symmetries in skyrmionic micromagnetic simulations: Cylindrical
  and radial meshes},}\ }\href {https://doi.org/10.1016/j.jmmm.2021.168972}
  {\bibfield  {journal} {\bibinfo  {journal} {Journal of Magnetism and Magnetic
  Materials}\ }\textbf {\bibinfo {volume} {549}},\ \bibinfo {pages} {168972}
  (\bibinfo {year} {2022})}\BibitemShut {NoStop}%
\bibitem [{\citenamefont {Yan}, \citenamefont {Wang},\ and\ \citenamefont
  {Campbell}(2008)}]{Yan_2008}%
  \BibitemOpen
  \bibfield  {author} {\bibinfo {author} {\bibfnamefont {M.}~\bibnamefont
  {Yan}}, \bibinfo {author} {\bibfnamefont {H.}~\bibnamefont {Wang}},\ and\
  \bibinfo {author} {\bibfnamefont {C.}~\bibnamefont {Campbell}},\ }\bibfield
  {title} {\enquote {\bibinfo {title} {Unconventional magnetic vortex
  structures observed in micromagnetic simulations},}\ }\href
  {https://doi.org/10.1016/j.jmmm.2008.02.170} {\bibfield  {journal} {\bibinfo
  {journal} {Journal of Magnetism and Magnetic Materials}\ }\textbf {\bibinfo
  {volume} {320}},\ \bibinfo {pages} {1937--1944} (\bibinfo {year}
  {2008})}\BibitemShut {NoStop}%
\bibitem [{\citenamefont {Karakas}\ \emph {et~al.}(2018)\citenamefont
  {Karakas}, \citenamefont {Gokce}, \citenamefont {Habiboglu}, \citenamefont
  {Arpaci}, \citenamefont {Ozbozduman}, \citenamefont {Cinar}, \citenamefont
  {Yanik}, \citenamefont {Tomasello}, \citenamefont {Tacchi}, \citenamefont
  {Siracusano}, \citenamefont {Carpentieri}, \citenamefont {Finocchio},
  \citenamefont {Hauet},\ and\ \citenamefont {Ozatay}}]{Karakas_2018}%
  \BibitemOpen
  \bibfield  {author} {\bibinfo {author} {\bibfnamefont {V.}~\bibnamefont
  {Karakas}}, \bibinfo {author} {\bibfnamefont {A.}~\bibnamefont {Gokce}},
  \bibinfo {author} {\bibfnamefont {A.~T.}\ \bibnamefont {Habiboglu}}, \bibinfo
  {author} {\bibfnamefont {S.}~\bibnamefont {Arpaci}}, \bibinfo {author}
  {\bibfnamefont {K.}~\bibnamefont {Ozbozduman}}, \bibinfo {author}
  {\bibfnamefont {I.}~\bibnamefont {Cinar}}, \bibinfo {author} {\bibfnamefont
  {C.}~\bibnamefont {Yanik}}, \bibinfo {author} {\bibfnamefont
  {R.}~\bibnamefont {Tomasello}}, \bibinfo {author} {\bibfnamefont
  {S.}~\bibnamefont {Tacchi}}, \bibinfo {author} {\bibfnamefont
  {G.}~\bibnamefont {Siracusano}}, \bibinfo {author} {\bibfnamefont
  {M.}~\bibnamefont {Carpentieri}}, \bibinfo {author} {\bibfnamefont
  {G.}~\bibnamefont {Finocchio}}, \bibinfo {author} {\bibfnamefont
  {T.}~\bibnamefont {Hauet}},\ and\ \bibinfo {author} {\bibfnamefont
  {O.}~\bibnamefont {Ozatay}},\ }\bibfield  {title} {\enquote {\bibinfo {title}
  {Observation of magnetic radial vortex nucleation in a multilayer stack with
  tunable anisotropy},}\ }\href {https://doi.org/10.1038/s41598-018-25392-x}
  {\bibfield  {journal} {\bibinfo  {journal} {Scientific Reports}\ }\textbf
  {\bibinfo {volume} {8}} (\bibinfo {year} {2018}),\
  10.1038/s41598-018-25392-x}\BibitemShut {NoStop}%
\bibitem [{\citenamefont {Lin}, \citenamefont {Reichhardt},\ and\ \citenamefont
  {Saxena}(2013)}]{Lin_2013}%
  \BibitemOpen
  \bibfield  {author} {\bibinfo {author} {\bibfnamefont {S.-Z.}\ \bibnamefont
  {Lin}}, \bibinfo {author} {\bibfnamefont {C.}~\bibnamefont {Reichhardt}},\
  and\ \bibinfo {author} {\bibfnamefont {A.}~\bibnamefont {Saxena}},\
  }\bibfield  {title} {\enquote {\bibinfo {title} {Manipulation of skyrmions in
  nanodisks with a current pulse and skyrmion rectifier},}\ }\href
  {https://doi.org/10.1063/1.4809751} {\bibfield  {journal} {\bibinfo
  {journal} {Applied Physics Letters}\ }\textbf {\bibinfo {volume} {102}},\
  \bibinfo {pages} {222405} (\bibinfo {year} {2013})}\BibitemShut {NoStop}%
\bibitem [{\citenamefont {Guslienko}\ \emph {et~al.}(2001)\citenamefont
  {Guslienko}, \citenamefont {Novosad}, \citenamefont {Otani}, \citenamefont
  {Shima},\ and\ \citenamefont {Fukamichi}}]{Guslienko_2001}%
  \BibitemOpen
  \bibfield  {author} {\bibinfo {author} {\bibfnamefont {K.~Y.}\ \bibnamefont
  {Guslienko}}, \bibinfo {author} {\bibfnamefont {V.}~\bibnamefont {Novosad}},
  \bibinfo {author} {\bibfnamefont {Y.}~\bibnamefont {Otani}}, \bibinfo
  {author} {\bibfnamefont {H.}~\bibnamefont {Shima}},\ and\ \bibinfo {author}
  {\bibfnamefont {K.}~\bibnamefont {Fukamichi}},\ }\bibfield  {title} {\enquote
  {\bibinfo {title} {Magnetization reversal due to vortex nucleation,
  displacement, and annihilation in submicron ferromagnetic dot arrays},}\
  }\href {https://doi.org/10.1103/PhysRevB.65.024414} {\bibfield  {journal}
  {\bibinfo  {journal} {Physical Review B}\ }\textbf {\bibinfo {volume} {65}},\
  \bibinfo {pages} {024414} (\bibinfo {year} {2001})}\BibitemShut {NoStop}%
\bibitem [{\citenamefont {Rana}\ and\ \citenamefont {Otani}(2019)}]{Rana_2019}%
  \BibitemOpen
  \bibfield  {author} {\bibinfo {author} {\bibfnamefont {B.}~\bibnamefont
  {Rana}}\ and\ \bibinfo {author} {\bibfnamefont {Y.}~\bibnamefont {Otani}},\
  }\bibfield  {title} {\enquote {\bibinfo {title} {Towards magnonic devices
  based on voltage-controlled magnetic anisotropy},}\ }\href
  {https://doi.org/10.1038/s42005-019-0189-6} {\bibfield  {journal} {\bibinfo
  {journal} {Communications Physics}\ }\textbf {\bibinfo {volume} {2}}
  (\bibinfo {year} {2019}),\ 10.1038/s42005-019-0189-6}\BibitemShut {NoStop}%
\bibitem [{\citenamefont {Suwardy}\ \emph {et~al.}(2019)\citenamefont
  {Suwardy}, \citenamefont {Goto}, \citenamefont {Suzuki},\ and\ \citenamefont
  {Miwa}}]{Suwardy_2019}%
  \BibitemOpen
  \bibfield  {author} {\bibinfo {author} {\bibfnamefont {J.}~\bibnamefont
  {Suwardy}}, \bibinfo {author} {\bibfnamefont {M.}~\bibnamefont {Goto}},
  \bibinfo {author} {\bibfnamefont {Y.}~\bibnamefont {Suzuki}},\ and\ \bibinfo
  {author} {\bibfnamefont {S.}~\bibnamefont {Miwa}},\ }\bibfield  {title}
  {\enquote {\bibinfo {title} {Voltage-controlled magnetic anisotropy and
  dzyaloshinskii-moriya interactions in {CoNi}/{MgO} and {CoNi}/pd/{MgO}},}\
  }\href {https://doi.org/10.7567/1347-4065/ab21a6} {\bibfield  {journal}
  {\bibinfo  {journal} {Japanese Journal of Applied Physics}\ }\textbf
  {\bibinfo {volume} {58}},\ \bibinfo {pages} {060917} (\bibinfo {year}
  {2019})}\BibitemShut {NoStop}%
\bibitem [{\citenamefont {Leon}\ \emph {et~al.}(2019)\citenamefont {Leon},
  \citenamefont {d{\textquotesingle}Albuquerque~e Castro}, \citenamefont
  {Retamal}, \citenamefont {Cahaya},\ and\ \citenamefont {Altbir}}]{Leon_2019}%
  \BibitemOpen
  \bibfield  {author} {\bibinfo {author} {\bibfnamefont {A.~O.}\ \bibnamefont
  {Leon}}, \bibinfo {author} {\bibfnamefont {J.}~\bibnamefont
  {d{\textquotesingle}Albuquerque~e Castro}}, \bibinfo {author} {\bibfnamefont
  {J.~C.}\ \bibnamefont {Retamal}}, \bibinfo {author} {\bibfnamefont {A.~B.}\
  \bibnamefont {Cahaya}},\ and\ \bibinfo {author} {\bibfnamefont
  {D.}~\bibnamefont {Altbir}},\ }\bibfield  {title} {\enquote {\bibinfo {title}
  {Manipulation of the {RKKY} exchange by voltages},}\ }\href
  {https://doi.org/10.1103/PhysRevB.100.014403} {\bibfield  {journal} {\bibinfo
   {journal} {Physical Review B}\ }\textbf {\bibinfo {volume} {100}},\ \bibinfo
  {pages} {014403} (\bibinfo {year} {2019})}\BibitemShut {NoStop}%
\bibitem [{\citenamefont {Hellwig}\ \emph {et~al.}(2007)\citenamefont
  {Hellwig}, \citenamefont {Berger}, \citenamefont {Kortright},\ and\
  \citenamefont {Fullerton}}]{Hellwig_2007}%
  \BibitemOpen
  \bibfield  {author} {\bibinfo {author} {\bibfnamefont {O.}~\bibnamefont
  {Hellwig}}, \bibinfo {author} {\bibfnamefont {A.}~\bibnamefont {Berger}},
  \bibinfo {author} {\bibfnamefont {J.~B.}\ \bibnamefont {Kortright}},\ and\
  \bibinfo {author} {\bibfnamefont {E.~E.}\ \bibnamefont {Fullerton}},\
  }\bibfield  {title} {\enquote {\bibinfo {title} {Domain structure and
  magnetization reversal of antiferromagnetically coupled perpendicular
  anisotropy films},}\ }\href {https://doi.org/10.1016/j.jmmm.2007.04.035}
  {\bibfield  {journal} {\bibinfo  {journal} {Journal of Magnetism and Magnetic
  Materials}\ }\textbf {\bibinfo {volume} {319}},\ \bibinfo {pages} {13--55}
  (\bibinfo {year} {2007})}\BibitemShut {NoStop}%
\bibitem [{\citenamefont {Ponsudana}\ \emph {et~al.}(2022)\citenamefont
  {Ponsudana}, \citenamefont {Amuda}, \citenamefont {Brinda},\ and\
  \citenamefont {Kanimozhi}}]{Ponsudana_2022}%
  \BibitemOpen
  \bibfield  {author} {\bibinfo {author} {\bibfnamefont {M.}~\bibnamefont
  {Ponsudana}}, \bibinfo {author} {\bibfnamefont {R.}~\bibnamefont {Amuda}},
  \bibinfo {author} {\bibfnamefont {A.}~\bibnamefont {Brinda}},\ and\ \bibinfo
  {author} {\bibfnamefont {N.}~\bibnamefont {Kanimozhi}},\ }\bibfield  {title}
  {\enquote {\bibinfo {title} {Investigation on the excitation of magnetic
  skyrmionium in a nanostructure},}\ }\href
  {https://doi.org/10.1007/s10948-021-06111-6} {\bibfield  {journal} {\bibinfo
  {journal} {Journal of Superconductivity and Novel Magnetism}\ } (\bibinfo
  {year} {2022}),\ 10.1007/s10948-021-06111-6}\BibitemShut {NoStop}%
\bibitem [{\citenamefont {Tejo}\ \emph
  {et~al.}(2020{\natexlab{a}})\citenamefont {Tejo}, \citenamefont {Saavedra},
  \citenamefont {Denardin},\ and\ \citenamefont {Escrig}}]{Tejo_2020}%
  \BibitemOpen
  \bibfield  {author} {\bibinfo {author} {\bibfnamefont {F.}~\bibnamefont
  {Tejo}}, \bibinfo {author} {\bibfnamefont {E.}~\bibnamefont {Saavedra}},
  \bibinfo {author} {\bibfnamefont {J.~C.}\ \bibnamefont {Denardin}},\ and\
  \bibinfo {author} {\bibfnamefont {J.}~\bibnamefont {Escrig}},\ }\bibfield
  {title} {\enquote {\bibinfo {title} {Dynamic susceptibility of skyrmion
  clusters in {Co/Pt} nanodots},}\ }\href {https://doi.org/10.1063/5.0023613}
  {\bibfield  {journal} {\bibinfo  {journal} {Applied Physics Letters}\
  }\textbf {\bibinfo {volume} {117}},\ \bibinfo {pages} {152401} (\bibinfo
  {year} {2020}{\natexlab{a}})}\BibitemShut {NoStop}%
\bibitem [{\citenamefont {Tejo}\ \emph
  {et~al.}(2020{\natexlab{b}})\citenamefont {Tejo}, \citenamefont {Velozo},
  \citenamefont {El{\'{\i}}as},\ and\ \citenamefont {Escrig}}]{Tejo_2020a}%
  \BibitemOpen
  \bibfield  {author} {\bibinfo {author} {\bibfnamefont {F.}~\bibnamefont
  {Tejo}}, \bibinfo {author} {\bibfnamefont {F.}~\bibnamefont {Velozo}},
  \bibinfo {author} {\bibfnamefont {R.~G.}\ \bibnamefont {El{\'{\i}}as}},\ and\
  \bibinfo {author} {\bibfnamefont {J.}~\bibnamefont {Escrig}},\ }\bibfield
  {title} {\enquote {\bibinfo {title} {Oscillations of skyrmion clusters in
  {Co/Pt} multilayer nanodots},}\ }\href
  {https://doi.org/10.1038/s41598-020-73458-6} {\bibfield  {journal} {\bibinfo
  {journal} {Scientific Reports}\ }\textbf {\bibinfo {volume} {10}} (\bibinfo
  {year} {2020}{\natexlab{b}}),\ 10.1038/s41598-020-73458-6}\BibitemShut
  {NoStop}%
\bibitem [{\citenamefont {Saavedra}\ \emph {et~al.}(2021)\citenamefont
  {Saavedra}, \citenamefont {Tejo}, \citenamefont {Vidal-Silva},\ and\
  \citenamefont {Escrig}}]{Saavedra_2021}%
  \BibitemOpen
  \bibfield  {author} {\bibinfo {author} {\bibfnamefont {E.}~\bibnamefont
  {Saavedra}}, \bibinfo {author} {\bibfnamefont {F.}~\bibnamefont {Tejo}},
  \bibinfo {author} {\bibfnamefont {N.}~\bibnamefont {Vidal-Silva}},\ and\
  \bibinfo {author} {\bibfnamefont {J.}~\bibnamefont {Escrig}},\ }\bibfield
  {title} {\enquote {\bibinfo {title} {Magnonic key based on skyrmion
  clusters},}\ }\href {https://doi.org/10.1038/s41598-021-02285-0} {\bibfield
  {journal} {\bibinfo  {journal} {Scientific Reports}\ }\textbf {\bibinfo
  {volume} {11}} (\bibinfo {year} {2021}),\
  10.1038/s41598-021-02285-0}\BibitemShut {NoStop}%
\bibitem [{\citenamefont {Kravchuk}\ \emph {et~al.}(2019)\citenamefont
  {Kravchuk}, \citenamefont {Gomonay}, \citenamefont {Sheka}, \citenamefont
  {Rodrigues}, \citenamefont {Everschor-Sitte}, \citenamefont {Sinova},
  \citenamefont {van~den Brink},\ and\ \citenamefont
  {Gaididei}}]{Kravchuk2019}%
  \BibitemOpen
  \bibfield  {author} {\bibinfo {author} {\bibfnamefont {V.~P.}\ \bibnamefont
  {Kravchuk}}, \bibinfo {author} {\bibfnamefont {O.}~\bibnamefont {Gomonay}},
  \bibinfo {author} {\bibfnamefont {D.~D.}\ \bibnamefont {Sheka}}, \bibinfo
  {author} {\bibfnamefont {D.~R.}\ \bibnamefont {Rodrigues}}, \bibinfo {author}
  {\bibfnamefont {K.}~\bibnamefont {Everschor-Sitte}}, \bibinfo {author}
  {\bibfnamefont {J.}~\bibnamefont {Sinova}}, \bibinfo {author} {\bibfnamefont
  {J.}~\bibnamefont {van~den Brink}},\ and\ \bibinfo {author} {\bibfnamefont
  {Y.}~\bibnamefont {Gaididei}},\ }\bibfield  {title} {\enquote {\bibinfo
  {title} {Spin eigenexcitations of an antiferromagnetic skyrmion},}\ }\href
  {https://doi.org/10.1103/PhysRevB.99.184429} {\bibfield  {journal} {\bibinfo
  {journal} {Physical Review B}\ }\textbf {\bibinfo {volume} {99}},\ \bibinfo
  {pages} {184429} (\bibinfo {year} {2019})}\BibitemShut {NoStop}%
\bibitem [{\citenamefont {Evans}\ \emph {et~al.}(2014)\citenamefont {Evans},
  \citenamefont {Fan}, \citenamefont {Chureemart}, \citenamefont {Ostler},
  \citenamefont {Ellis},\ and\ \citenamefont {Chantrell}}]{Evans2014}%
  \BibitemOpen
  \bibfield  {author} {\bibinfo {author} {\bibfnamefont {R.~F.~L.}\
  \bibnamefont {Evans}}, \bibinfo {author} {\bibfnamefont {W.~J.}\ \bibnamefont
  {Fan}}, \bibinfo {author} {\bibfnamefont {P.}~\bibnamefont {Chureemart}},
  \bibinfo {author} {\bibfnamefont {T.~A.}\ \bibnamefont {Ostler}}, \bibinfo
  {author} {\bibfnamefont {M.~O.~A.}\ \bibnamefont {Ellis}},\ and\ \bibinfo
  {author} {\bibfnamefont {R.~W.}\ \bibnamefont {Chantrell}},\ }\bibfield
  {title} {\enquote {\bibinfo {title} {Atomistic spin model simulations of
  magnetic nanomaterials},}\ }\href
  {https://doi.org/10.1088/0953-8984/26/10/103202} {\bibfield  {journal}
  {\bibinfo  {journal} {Journal of Physics: Condensed Matter}\ }\textbf
  {\bibinfo {volume} {26}},\ \bibinfo {pages} {103202} (\bibinfo {year}
  {2014})}\BibitemShut {NoStop}%
\end{thebibliography}%


\clearpage

\end{document}